\documentclass[a4paper,usenatbib]{mnras}
\usepackage{aas_macros}
\usepackage{ae,aecompl}
\usepackage{graphicx}
\usepackage{amsmath}
\usepackage{amssymb}
\usepackage{color}
\usepackage{microtype}

\newcommand{\Msun}{\ensuremath{M_{\odot}}}

\newcommand{\hMpc}{\ensuremath{h^{-1}{\rm Mpc}}}

\newcommand{\avg}[1]{\ensuremath{\left< #1 \right>}}

\newcommand{\der}{\ensuremath{{\rm d}}}

\newcommand{\erfc}[1]{\ensuremath{{\rm erfc}\left(#1\right)}}
\newcommand{\erf}[1]{\ensuremath{{\rm erf}\left(#1\right)}}

\newcommand{\eqn}[1]{equation~\eqref{#1}}
\newcommand{\eqns}[1]{equations~\eqref{#1}}

\newcommand{\be}{\begin{equation}}
\newcommand{\ee}{\end{equation}}

\newcommand{\ph}[1]{\phantom{#1}}

\newcommand{\Ngal}{\ensuremath{N}}

\newcommand{\ngal}{\ensuremath{\bar n_{\rm g}}}

\newcommand{\Ns}{\ensuremath{\bar N_{\rm sat}}}

\newcommand{\Nsnew}{\ensuremath{\bar N_{\rm sat}^{\rm new}}}

\newcommand{\fcen}{\ensuremath{f_{\rm cen}}}
\newcommand{\fcennew}{\ensuremath{f_{\rm cen}^{\rm new}}}

\newcommand{\Lmin}{\ensuremath{L_{\rm min}}}

\definecolor{darkgreen}{rgb}{0,0.6,0}
\definecolor{orange}{rgb}{1.0,0.5,0.0}

\title[Halo Model and Order Statistics]{An Order Statistics Approach to the Halo Model for Galaxies}

\author[N. Paul et al.]{
Niladri Paul,$^{1}$\thanks{E-mail: npaul@iucaa.in}
Aseem Paranjape$^{1}$\thanks{E-mail: aseem@iucaa.in}
and Ravi K. Sheth$^{2}$\thanks{E-mail: shethrk@sas.upenn.edu}
\\
$^{1}$Inter-University Centre for Astronomy and Astrophysics, Ganeshkhind, Post Bag 4, Pune 411007, India\\
$^{2}$Center for Particle Cosmology, University of Pennsylvania, 209 S. 33rd St., Philadelphia, PA 19104, USA\\
}

\date{draft}

\pubyear{2016}

\begin{document}
\label{firstpage}
\pagerange{\pageref{firstpage}--\pageref{lastpage}}
\maketitle

\begin{abstract} 
We use the Halo Model to explore the implications of assuming that galaxy luminosities in groups are randomly drawn from an underlying luminosity function. 
We show that even the simplest of such order statistics models -- one in which this luminosity function $p(L)$ is universal -- naturally produces a number of features associated with previous analyses based on the `central plus Poisson satellites' hypothesis. 
These include the monotonic relation of mean central luminosity with halo mass, the Lognormal distribution around this mean, and the tight relation between the central and satellite mass scales. 
In stark contrast to observations of galaxy clustering, however, this model predicts \emph{no} luminosity dependence of large scale clustering. 
We then show that an extended version of this model, based on the order statistics of a \emph{halo mass dependent} luminosity function $p(L|m)$, is in much better agreement with the clustering data as well as satellite luminosities, but systematically under-predicts central luminosities. 
This brings into focus the idea that central galaxies constitute a distinct population that is affected by different physical processes than are the satellites. 
We model this physical difference as a statistical brightening of the central luminosities, over and above the order statistics prediction. 
The magnitude gap between the brightest and second brightest group galaxy is predicted as a by-product, and is also in good agreement with observations. 
We propose that this order statistics framework provides a useful language in which to compare the Halo Model for galaxies with more physically motivated galaxy formation models. 
\end{abstract}

\begin{keywords}
galaxies: groups: general -- cosmology: large-scale structure of Universe -- methods: analytical 
\end{keywords}


\section{Introduction} \label{sec:introduction}
\noindent
The standard paradigm of galaxy formation in a hierarchically evolving Universe places galaxies in groups and clusters that reside in dark matter haloes. This Halo Model view of galaxies \citep[see][for a review]{Cooray_Sheth_2002} has found considerable observational support in recent years \citep[see, e.g.,][hereafter, Z11]{Zehavi_et_al_2011}. In particular, the Halo Model provides a language for describing the striking observation that the spatial clustering of galaxies is a strong function of luminosity, at least at $L>L_\ast$; it essentially does so by allowing one to determine how the galaxy population in a halo depends on halo mass. This determination is done in either of two ways. The Halo Occupation Distribution (HOD) approach determines how the mean number of galaxies above some $L$ depends on halo mass \citep{Zehavi_et_al_2005}. Repeating this for a range of threshold luminosities it leads to a characterisation of how the galaxy luminosity function depends on halo mass. Initial versions of this algorithm required the fitting of 3 free parameters for each threshold $L$ \citep{Zehavi_et_al_2005}; more recent versions, which distinguish explicitly between the central galaxy in a halo (which is typically the brightest) and all the others, require 5 \citep{Zehavi_et_al_2011,Guo_et_al_2015}. The Conditional Luminosity Function (CLF) approach, on the other hand, explicitly postulates a Lognormal shape for the central galaxy (2 halo-mass dependent parameters), and a Schechter-like function for the other galaxies (3 or 4 halo-mass dependent parameters) whose free parameters must be determined from simultaneously matching the abundance and clustering as a function of $L$ \citep{Yang_et_al_2008,Cacciato_2012}.  

This separate modelling of the central or brightest galaxy in a group -- often called the BGG (or BCG in literature dealing with galaxy clusters) -- is consistent with the fact that this class of objects appears to follow slightly different scaling laws than those defined by the bulk of the galaxy population \citep[see, e.g.,][]{Bernardi_et_al_2011}. Despite this dichotomy, the BGG \emph{luminosity function} is, in fact, reasonably consistent with the hypothesis that BGG luminosities are merely the statistical extremes of the group galaxy luminosity function. Anecdotal evidence that the match with extreme value statistics must be quite good comes from the fact that the issue is still debated \citep{Schechter_peebles_1976, Tremaine_&_Richstone_1977, Bhavsar_Barrow_1985, Loh_Strauss_2006, Vale_Ostriker_2008,Ostriker_Miller_2010, Dobos_Casabai_2011}, some fifty years after it was first raised by \cite{Scott_1957}. Recent work has shown that the luminosity function of the second brightest galaxy in a group is also in good agreement with the order statistics prediction \citep{Paranjape_Sheth_2012}.  

Relying purely on 1-point statistics has its own pitfalls, however; a point that was highlighted by \cite{Paranjape_Sheth_2012} using a marked correlation analysis which indicated that the spatial distribution of BGGs in SDSS groups is inconsistent with the assumptions of order statistics based on a universal luminosity function. Subsequent work used a different kind of 2-point statistic -- the magnitude gap between the first and second brightest group galaxies -- to reach similar conclusions \citep{Hearin_et_al_2012,Surhud_2012}. More recently, using updated data sets, \citet{Shen_et_al_2014} have reinforced this point by demonstrating that order statistics -- even allowing for an observationally constrained group richness dependence in the underlying luminosity function -- in fact fails to reproduce the BGG luminosity function in a systematic and statistically significant way.

It would therefore seem that the order statistics hypothesis for galaxy luminosities is strongly disfavoured by the data and should be discarded. We will argue below that, on the contrary, there is still much that can be learned from marrying order statistics to the Halo Model and studying the consequences, even if we know that the hypothesis in its original form does not describe the data. Firstly, we will show that the assumption of a universal galaxy luminosity function vastly simplifies Halo Model analyses. While it is almost trivial to see why this will be true for the CLF approach (the luminosity function is universal, so there is no longer a need to split the luminosity function up into central plus satellites), we will show why Order statistics provide considerable insight into previous CLF based results: e.g., the extreme value shape of the luminosity distribution of the brightest galaxy is reasonably well approximated by a Lognormal of approximately the right width. Perhaps more importantly, our analysis shows clearly just how much simpler the HOD analyses could be:  in particular, calibration of an HOD is required on the faintest luminosity threshold only. Although not a good description of the clustering data, as we will demonstrate, our analysis can still provide an excellent framework for the setting of initial guesses and priors that are part and parcel of modern HOD fitting routines.  

Secondly, we will explore whether the situation improves upon relaxing the criterion of universality and allowing for a halo mass dependence in the luminosity function that forms the basis of the order statistics hypothesis. This is, in a sense, a revival of some of the ideas presented by \citet{Vale_Ostriker_2006, Vale_Ostriker_2008}, now using statistics at fixed halo mass \emph{and} group richness. Since order statistics no longer predict how the luminosity function should depend on halo mass, this model does not reduce the complexity of the HOD analysis as much as if the luminosity function were universal. And while the resulting model can be brought into substantially better agreement with the data, it systematically predicts fainter BGGs than are observed.

The final piece in the puzzle, as we will argue, is to model an additional brightening of the central galaxy, over and above what is predicted by order statistics \citep[c.f.][]{Shen_et_al_2014}. This not only brings the model into good agreement with a wide range of observables, but also isolates the physics that makes the centrals special by modelling it as an extra brightening. The statistics of the magnitude gap are then a prediction of this model, and are also in reasonable agreement with the data.

The paper is organised as follows. In section~\ref{sec:extreme_value_mass_independent} we describe the analytical framework for combining the halo model with order statistics based on a universal luminosity function, and its observational consequences. In section~\ref{sec:extreme_value_mass_dependent}, we incorporate halo mass dependence in the model and in section~\ref{sec:centrals_convolved}, we present our final model which brightens the centrals over and above the order statistics prediction. We discuss the predicted magnitude gap statistics from all these models in section~\ref{sec:gap}, and conclude in section~\ref{sec:conclude}. 
	
Throughout this work, we use a flat Lambda-cold dark matter ($\Lambda$CDM) cosmology with matter density parameter $\Omega_m = 0.25$, baryon density parameter $\Omega_b = 0.045$, Hubble parameter $H_0 = 100 h$ km $\rm s^{-1}$ $\rm Mpc^{-1}$ with $h = 0.7$, primordial r.m.s. density fluctuations at the scale of $8 h^{-1}$Mpc, $\sigma_8 = 0.8$ and an inflationary spectral index, $n_s = 0.95$, consistent with the 5-year results of the Wilkinson Microwave Anisotropy Probe experiment \citep{Komatsu_et_al_2009_WMAP5}, and also with the values assumed by \citet{Zehavi_et_al_2011} in calibrating their HOD, which we will use below. Wherever needed, we have used the \cite{Eisenstein_Hu_1999} fitting function for the linear theory matter power spectrum. We will quote halo masses in $h^{-1}M_\odot$ and luminosities using SDSS $r$-band absolute magnitudes, $K$-corrected and evolution corrected to $z=0.1$ and denoted $M_{{}^{0.1}r}$ \citep{Blanton_et_al_2003}, always quoting values of $M_r \equiv M_{{}^{0.1}r} - 5\log_{10}(h)$.

\section{The Halo Model with a universal luminosity function} 
\label{sec:extreme_value_mass_independent}
\noindent
The HOD fitting procedure \citep{Berlind_weinberg_2002, Zehavi_et_al_2005} assumes that only a fraction $f(m)$ of haloes of mass $m$ host at least one galaxy brighter than some luminosity threshold $L_{\rm min}$. Typically $f(m)$ increases monotonically, from 0 to 1, with $m$. Since (massive) haloes may host many galaxies brighter than $L_{\rm min}$, the HOD procedure adds a second term -- typically one that varies as a power law in $m$ -- to account for the possibility that the mean number is larger than unity.  For reasons which will become clear shortly, the term $f$ is often associated with the first, brightest galaxy in the halo.  Typically, it is assumed that this first galaxy sits at the halo center (though this is not necessary for the 
formalism); this gives rise to suggestive jargon in which the other objects are called satellites.  If one assumes that only haloes with centrals can host satellites, then the HOD description boils down to the assumption that the mean number of galaxies in a halo of mass $m$ is 
\begin{equation}
\avg{N|m} = f_{\rm cen}(m)\,[1 + \bar N_{\rm sat}(m)].
\label{meanNLmin}
\end{equation}
To describe clustering statistics, the HOD prescription actually requires slightly more information:  For $n$-point statistics, it requires the $n$-th moment of $p(N|m)$, the distribution of the number of galaxies in haloes of mass $m$.  In principle, this requires the specification of a potentially large number of free parameters.  In practice, the assumption that, in haloes which host a central, the satellites follow a Poisson distribution with mean $\bar N_{\rm sat}(m)$ (so that all $n$-point statistics are fully specified without having to fit any additional free parameters), has proved to be remarkably accurate. This central plus Poisson satellites model is typically fit to clustering measurements for a range of threshold $L$ values.  This means that $f_{\rm cen}(m)$ and $\bar N_{\rm sat}(m)$ depend on this threshold $L$:  i.e., the free parameters of these functional forms must be refit for each threshold $L$. As we show below, the assumption that the galaxy luminosity function is universal above some $L_{\rm min}$ means that one need only determine $f_{\rm cen}(m)$ and $\bar N_{\rm sat}(m)$ for $L_{\rm min}$ -- there is no need to perform further fits. Alternatively, performing HOD fits to samples with brighter thresholds provides a test of the assumption of universality. 

\begin{figure}
 \centering
 \includegraphics[width=0.45\textwidth]{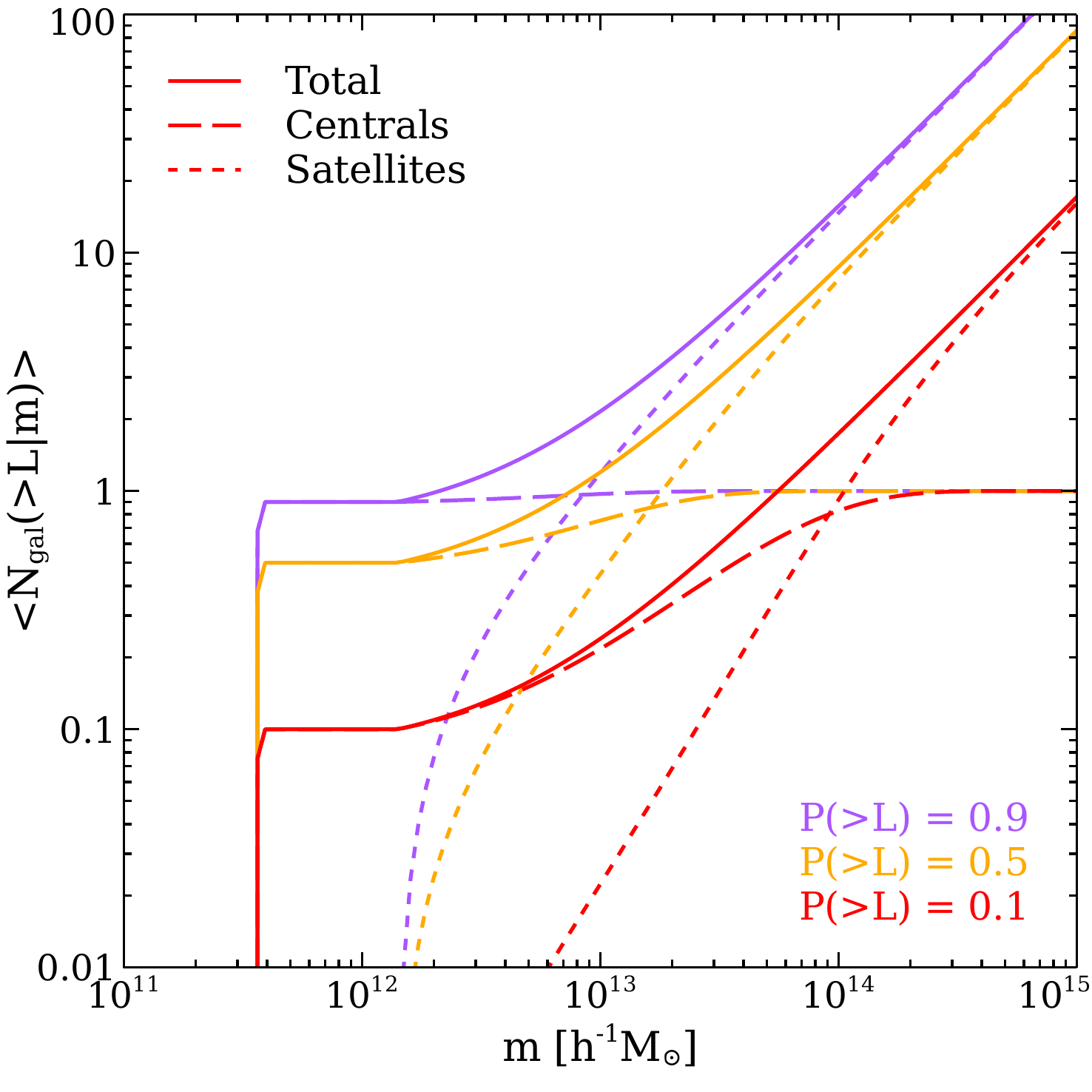}
 \caption{Predicted $\avg{N|m}$ for $p(>L) = (0.9,0.5,0.1)$ from top to bottom, 
          where the starting relation for $L_{\rm min}$ was taken to be 
          that for objects with $M_r<-19.5$ in the SDSS (from Table~3 of 
          Zehavi et al. 2011).  Solid curves show \eqn{meanNL}, and
          can be thought of as the sum of the dashed 
          and dotted curves, which represent the central and satellite 
          contributions as in \eqn{meanNL-censat}.  These show many of
          the same trends seen in Figure~10 of \citet{Zehavi_et_al_2011}.}
\label{fig-xHOD}
\end{figure}

We start by setting up the analytical predictions of universality for the HOD, emphasizing that this sole requirement
correctly reproduces many of the trends seen in current data and HOD analyses. In the following subsection we perform a quantitative comparison of the new framework with the existing HOD results, discussing its shortcomings.

\subsection{Analytical framework}
\noindent
\subsubsection{Luminosity functions}
\noindent
Suppose that the galaxy luminosity function is universal above some $L_{\rm min}$.  By universal, we mean that the shape of the galaxy luminosity function in a group is independent of the number of galaxies in the group -- only its normalization changes. In other words, the fraction $p(>L)$ of galaxies brighter than some $L>L_{\rm min}$ is the same for
groups containing different numbers of galaxies. This appears to be a good approximation for $L_{\rm min}\sim L_*$ 
\citep{Hansen_et_al_2009,Paranjape_Sheth_2012}.  

If \eqn{meanNLmin} holds for the threshold $L_{\rm min}$, this immediately implies that the mean number of galaxies brighter than some new threshold $L>L_{\rm min}$ in haloes of mass $m$, is simply 
\be
\avg{N|>L,m} = f_{\rm cen}(m)\,[1 + \bar N_{\rm sat}(m)]\,p(>L)\,. 
\label{meanNL}
\ee
Figure~\ref{fig-xHOD} shows the trends predicted by this requirement.

\begin{figure}
 \centering
 \includegraphics[width=0.45\textwidth]{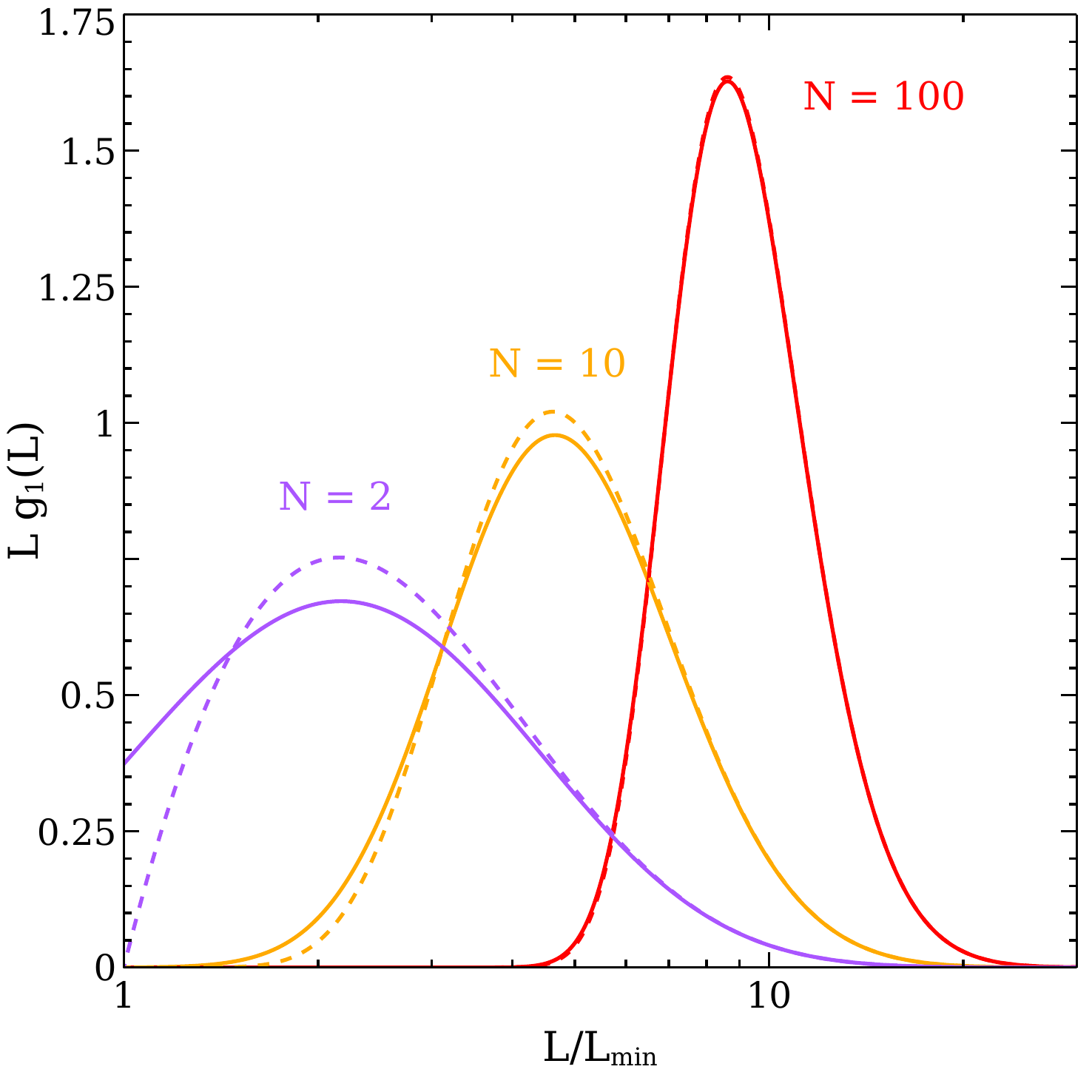}
 \caption{Predicted differential luminosity distribution of the most luminous galaxy 
          in objects which have $N$ galaxies brighter than $L_{\rm min}$, with corresponding cumulative distribution  
          $g_1(<L)=p(<L)^{N}$ (dotted), and in
          objects of mass $m$ which have a Poisson distribution of
          satellites brighter than $L_{\rm min}$ (solid, with the cumulative distribution of  
          equation~\ref{p0L}), with mean $\Ns$ for $N=2, 10$ and
          $100$, and an underlying luminosity function 
          $p(>L)$ obtained from the Schechter function fit of \citet{Blanton_et_al_2003}. Using HOD
          parameters appropriate for $M_r<-19.5$ from Table 3 of \citet{Zehavi_et_al_2011}, 
          the three solid curves correspond,
          respectively, to $\log [m/h^{-1}\Msun] =12.9, 13.7$ and
          $14.8$.}
\label{fig-lg1l}
\end{figure}

Further, notice that the quantity $f_{\rm cen}(m)$ is really counting  the fraction of haloes of mass $m$ which have at least one galaxy brighter than $L_{\rm min}$.  If we increase the threshold luminosity, then this fraction will decrease:  we will use $f_{\rm cen}(>L|m)$ to denote this new value.  The assumption of a universal luminosity function allows us to quantify $f_{\rm cen}(>L|m)$ as follows.  Suppose the halo had $N_s+1$ galaxies brighter than $L_{\rm min}$.  The probability that none of these was brighter than the new threshold $L>L_{\rm min}$ is $p(<L)^{N_{\rm s} + 1}$.  Averaging this over $N_{\rm s}$ yields the probability $p_0(>L|m)$ that there are no galaxies brighter than $L$, in haloes containing at least one galaxy brighter than $L_{\rm min}$,
\begin{align}
 p_0(>L|m) &= \sum_{N_{\rm s}\ge 0} 
    \frac{\bar N_{\rm sat}(m)^{N_{\rm s}}{\rm e}^{-\bar N_{\rm
          sat}(m)}}{N_{\rm s}!} \,p(<L)^{N_{\rm s} + 1}\nonumber\\
          &= p(<L)\,{\rm e}^{-\bar N_{\rm sat}(m)p(>L)}.
 \label{p0L}
\end{align}
Therefore, 
\begin{equation}
 f_{\rm cen}(>L|m) = f_{\rm cen}(m)\Bigl[1 - p_0(>L|m)\Bigr]
 \label{fcenL}
\end{equation} 
We stated earlier that this quantity is assumed to encode information about the BGG luminosity function.  It is particularly easy to see why in the current context.  This is because the probability $p_0(>L|m)$ that no galaxies are brighter than $L$ equals the probability $g_1(<L|m)$ that the brightest galaxy is fainter than $L$. Therefore, differentiating $f_{\rm cen}(>L|m)$ with respect to $L$ yields a quantity which is proportional to the shape of the
distribution of BGG luminosities in haloes of mass $m$. In particular, at large $m$, where $f_{\rm cen}(m)\to 1$,  $f_{\rm cen}(>L|m) \to 1 - p_0(>L|m) = g_1(>L|m)$ is a direct measure of the bright end of the BGG luminosity function. Figure~\ref{fig-lg1l} compares the BGG luminosity distributions in haloes with fixed mass $m$ to those in haloes with a fixed number of galaxies $N$. At large $m$ or $N$, the difference is small, and one correctly recovers observed trends such as an approximately Lognormal shape with a width that decreases and a mean that increases weakly with $m$ or $N$ (see below). 

The previous analysis also leads, in a very simple manner, to the expression for the mean number of satellites brighter than $L$: one can decompose the expression for $\avg{N|>L,m}$ into the contribution from centrals and satellites as
\begin{equation}
 \avg{N|>L,m} =
  f_{\rm cen}(>L|m) + f_{\rm cen}(m)\bar N_{\rm sat}(>L|m)\,,
\label{meanNL-censat}
\end{equation}
where $\bar N_{\rm sat}(>L|m)$ is the mean number of satellites brighter than $L$, in haloes that have at least one galaxy brighter than $L_{\rm min}$. Equations~\eqref{meanNL} and~\eqref{fcenL} lead to
\begin{align}
 \bar N_{\rm sat}(>L|m)
 &= \bar N_{\rm sat}(m)\,p(>L) \nonumber\\
 &\ph{\,p(>L) }
 - p(<L) \left[1 - {\rm e}^{-\bar N_{\rm sat}(m) p(>L)}\right]\nonumber\\
   &= [1 + \bar N_{\rm sat}(m)]\,p(>L) - \frac{f_{\rm cen}(>L|m)}{f_{\rm cen}(m)}.
\label{meanNsatL}
\end{align}
Alternatively, one can also compute $\bar N_{\rm sat}(>L|m)$ as the mean number of satellites brighter than $L$ in haloes containing exactly $N_s$ satellites brighter than $L_{\rm min}$, $N_sp_{\rm sat}(>L|N_s)$, averaged over the distribution of
$N_s$, 
\begin{equation}
 \bar N_{\rm sat}(>L|m) = \sum  p(N_s|m)\,N_s\,
                               p_{\rm sat}(>L|N_s)\,.
\end{equation}
The distribution $p_{\rm sat}(>L|N_s)$ follows from extreme value statistics \cite[see their equation A1]{Paranjape_Sheth_2012},
\begin{equation}
 p_{\rm sat}(>L|N_s) = p(>L) - \frac{p(<L) - p(<L)^{N_s + 1}}{N_s}\,, 
\label{eq:p_sat}
\end{equation}
and upon using the Poisson distribution for $p(N_s|m)$, one recovers \eqn{meanNsatL} for $\bar N_{\rm sat}(>L|m)$.

Figure~\ref{fig-xHOD} shows that the decomposition of $\avg{N|>L,m}$ into central and satellite terms (equation~\ref{meanNL-censat}) is rather similar to what HOD and CLF analyses routinely find, with the transition from one to the other becoming less pronounced as the threshold $L$ increases. A minor technical difference with the order statistics decomposition is that, in the standard HOD approach, one typically defines $\Ns(>L|m)$ to be the average number of satellites brighter than $L$ in a halo of mass $m$ that has a central of luminosity greater than $L$ (rather than $L_{\rm min}$), leading to
\begin{eqnarray}
\avg{\Ngal|>L,m} = \fcen(>L|m) \left[ 1 + \Ns(>L|m) \right] \,\, . 
\label{eq:Ng_avg_standard_HOD}
\end{eqnarray}
Finally, a result that will be useful later is the order statistics prediction for luminosity distribution of the $n^{\rm th}$ brightest galaxy in a group of $N$ galaxies, i.e., the distribution of the $n^{\rm th}$ largest of $N$ independent draws from an underlying distribution; for a universal luminosity distribution $p(L)$, this is given by
\be
g_n(L|N) = \binom{N}{n}\, n\, p(L)\,p(>L)^{N-n}\, p(<L)^{n-1} \,. 
\label{eq:differential_g_N_n}
\ee

\subsubsection{2-point statistics}
\noindent
Exactly like in the case of the usual HOD framework, the formalism developed above can be used in predicting the luminosity function and 2-point correlation function of galaxies, essentially by averaging the appropriate counting statistics over the distribution of haloes $n(m)$\footnote{$n(m)$ is the number density of haloes in the mass range $(m,m+\der m)$, i.e., the halo mass function. Throughout, we use the fitting form prescribed by  \cite{Tinker_2008_mass_function}.}. The number density of galaxies brighter than a threshold $L$ is
\be
\ngal (>L) = \int \der m\, n(m)\,\avg{\Ngal|>L,m} \,, 
\label{eq:avg_ng}
\ee
while the correlation function can be most easily expressed by splitting it into the so-called 2-halo and 1-halo pieces. The 2-halo term in Fourier space can be approximated as 
\begin{align}
P_{\rm 2h}(k) &= P_{\rm lin}(k) \bigg[\int \frac{\der m\,n(m)}{\ngal (>L)} b(m)\bigg\{  \fcen(>L|m) \notag\\
&\ph{P_{\rm lin}(k)\bigg[f\bigg]}  
+ \fcen(m) \Ns(>L|m)u(k|m) \bigg\} \bigg]^2 \,, 
\label{eq:P2h}
\end{align}
where $P_{\rm lin}(k)$ is the linear theory matter power spectrum and $b(m)$ is the linear halo bias\footnote{We use the expression for halo bias given by \cite{Tinker_2010_halo_bias}.}. $u(k|m)$ is the normalized Fourier transform of the profile $\rho(r|m)$ with which satellites are distributed around their central galaxy; this also defines the 1-halo term as follows: 
\begin{align}
\xi_{\rm 1h}(r) &= \int \frac{\der m\,n(m)}{\ngal (>L)^2}\bigg[ 2 \fcen(>L|m)\Ns(>L|m)\frac{\rho(r|m)}{m} \notag\\  
&\ph{\int\frac12\ngal}
+ \fcen(m)\,\Ns(>L|m)^2\frac{\lambda(r|m)}{m^2}  \bigg]\,,
\label{eq:xi_1h(r)}
\end{align} 
where $\lambda(r|m)$ is the convolution of $\rho(r|m)$ with itself\footnote{Throughout, we use an NFW profile \citep*{Navarro_1997} for $\rho(r|m)$, with a mean concentration-mass relation $\bar c(m)$ as used by \cite{Zehavi_et_al_2011} which itself is a slightly modified version of the relation given in \cite{Bullock_et_al_2001}. We do not introduce scatter around this mean relation.} and we have used the Poisson distribution for satellite counts to replace the second factorial moment $\avg{N_s(N_s-1)|>L,m}$ with $\Ns(>L|m)^2$ in the second line. The total correlation function in real space is then $\xi(r) = \xi_{\rm 1h}(r) + \xi_{\rm 2h}(r)$. 

In principle, in order to achieve $\sim5\%$ accuracy in the model, we should also include the effects of scale-dependent halo bias and halo exclusion in the 2-halo term, non-Poissonian effects in satellite pair counts and scatter in the concentration-mass relation in the 1-halo term, and the effects of redshift space distortions \citep[see, e.g.,][]{V_Bosch_et_al_2013}. Since our aim in this paper is to present a proof of principle rather than an optimized model, for simplicity we will ignore these additional effects.

\begin{figure}
\centering
\includegraphics[width=0.45\textwidth]{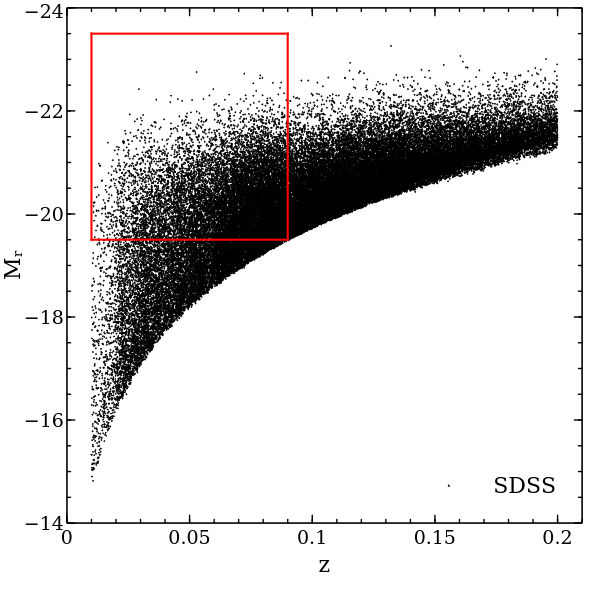}
\caption{SDSS DR7 galaxy sample, showing redshift against Petrosian $r$-band absolute magnitude $M_r$, K-corrected and evolution corrected to $z=0.1$. For all our analysis presented in this article, we use a volume limited sample of galaxies with $-23.5 < M_r \leq -19.5$ and $0.01 < z < 0.09$ as demarcated by the red box.} \label{fig:SDSS_data}
\end{figure}

The expressions for $P_{\rm 2h}$ and $\xi_{\rm 1h}$ above are valid only for the extreme value model that we are discussing. In the standard HOD framework, one must replace the term $\fcen(m)$ with $\fcen(>L|m)$ in these expressions, which is because the definition of the HOD is different in the two cases as is clear from \eqn{meanNL-censat} and \eqn{eq:Ng_avg_standard_HOD}. 

To mitigate redshift space effects, correlation measurements are often quoted using the projected correlation function $w_{\rm p}(r_{\rm p})$ given by \citep{Davis_Peebles_1983} 
\begin{align}
w_{\rm p}(r_{\rm p}) &= \int_0^\infty\der r_\pi\,\xi(r_{\rm p},r_\pi) = 2 \int_{r_{\rm p}}^{\infty}\frac{\der r\,r\,\xi(r)}{\sqrt{r^2 - r_{\rm p}^2}}\,,
\label{eq:wp_rp}
\end{align}
where $r_{\rm p}$ ($r_\pi$) is the separation between two galaxies perpendicular (parallel) to the line of sight. Note that all the quantities $\xi_{1h}(r)$, $\xi_{2h}(r)$ and $w_{\rm p}(r_{\rm p})$ depend on the luminosity threshold $L$ which we have omitted for brevity. 

\subsection{Comparison with observations}
\noindent
We now turn to some of the quantitative implications of this new HOD prescription. Many of our results below will rely on the group catalog of \citet{Yang_et_al_2007}\footnote{http://gax.shao.ac.cn/data/Group.html}. This catalog was constructed using the halo-based group finder described in \cite{yang+05} to identify groups in the New York University Value Added Galaxy Catalog \citep[NYU-VAGC;][]{blanton+05}, based on the Sloan Digital Sky Survey\footnote{http://www.sdss.org} \citep[SDSS;][]{york+00} data release 7 \citep[DR7;][]{abazajian+09}. Throughout, we will restrict attention to a volume limited subsample containing galaxies with spectroscopic redshifts \citep[`sample II' of][]{Yang_et_al_2007} restricted to the range $0.01 < z < 0.09$, with the Petrosian absolute magnitudes to the range $-23.5 < M_r < -19.5$. This is shown in Figure~\ref{fig:SDSS_data}.

\begin{figure}
 \centering
 \includegraphics[width=0.45\textwidth]{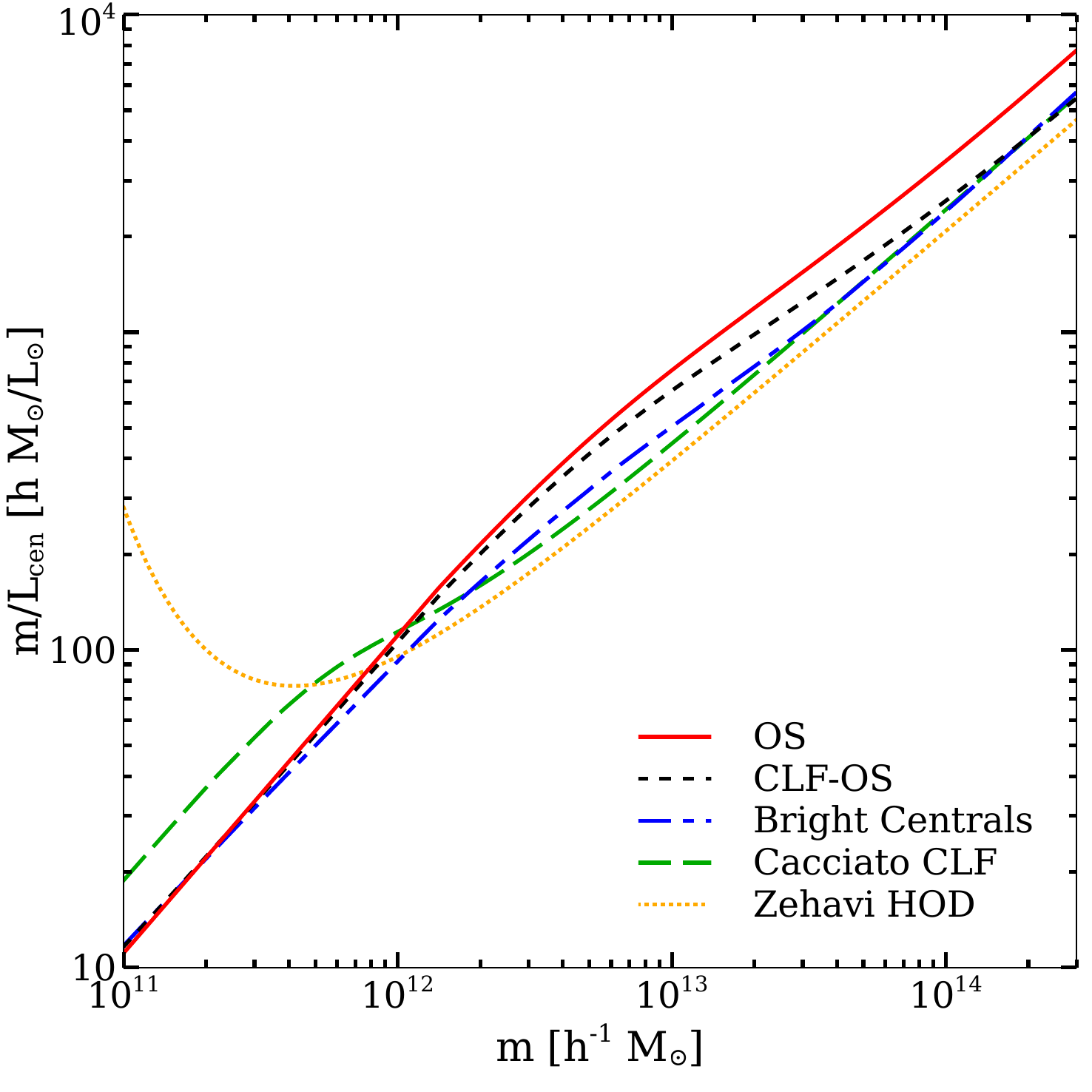}
 \caption{Ratio of halo mass to the median of the BGG luminosity
   function. For our analysis of the order statistics model with a universal luminosity function, the latter is calculated as the median
   of $p_0(>L|m)$, with the result shown as the solid red curve. The
   fine-dotted yellow curve shows the fit from \citet{Zehavi_et_al_2011}, in
   which $L_{\rm cen}$ corresponds to the median of $f_{\rm cen}(>L|m)$ (i.e., their $\avg{N_{\rm cen}(m,L)}$; see their
   equation~12 and Figure~12(b)), and the long-dashed green curve shows the result of using the standard (central+satellite) CLF model of \citet{Cacciato_2012}, with $L_{\rm cen}$ being the median of their $p_{\rm cen}(>L|m)=\phi_{\rm cen}(>L|m)/\phi_{\rm cen}(>L_{\rm min}|m)$. For comparison, the dotted black and dot-dashed blue curves show the results of our extended order statistics models described in sections~\ref{sec:extreme_value_mass_dependent} and~\ref{sec:centrals_convolved}, respectively. We see that all the order statistics models qualitatively predict the same monotonic trend as the standard CLF model.} 
\label{fig-mhbylcen}
\end{figure}

\subsubsection{The shape of the BGG luminosity function}
\noindent
Some HOD analyses associate the relation between $m$ and $L$ which comes from requiring $f_{\rm cen}(>L|m)=1/2$ with the mass-to-light ratio of the BGG.
The analogous relation between $m$ and $L$ in our analysis is given by $1/2 = p(<L)\,{\rm e}^{-\bar N_{\rm sat}(m_{1/2})p(>L)}$, which implies that 
\begin{equation}
 \bar N_{\rm sat}(m_{1/2})p(>L) = \ln[2p(<L)] \approx \ln(2)
\label{medianbcg}
\end{equation}
at large $L$.  If $\bar N_{\rm sat}(m)\propto m^\alpha$, and the luminosity function falls as $\exp(-L/L_*)$, then this says that BGG luminosity is only a weak function of mass at large $L$. Figure~\ref{fig-mhbylcen} shows the ratio $m/L_{\rm cen}$, where $L_{\rm cen}$ is the median of the BGG luminosity distribution. In our analysis, the latter is calculated as in \eqn{medianbcg} using the \citet{Blanton_et_al_2003} luminosity function, and the result is shown by the solid red curve. The fine-dotted yellow curve shows the fit of Z11, computed as the median of $f_{\rm cen}(>L|m)$ (i.e., their $\avg{N_{\rm cen}(m,L)}$; see their equation~12 and Figure~12(b)), while the long-dashed green curve shows the result of using the standard (central+satellite) CLF model of \citet{Cacciato_2012} for the central luminosities. For comparison, the other curves show the results of our extended models from sections~\ref{sec:extreme_value_mass_dependent} and~\ref{sec:centrals_convolved} below. We see that our universal order statistics prediction, as well as those from the extended models, are in good qualitative agreement with the standard CLF prediction.

\begin{figure}
 \centering
 \includegraphics[width=0.45\textwidth]{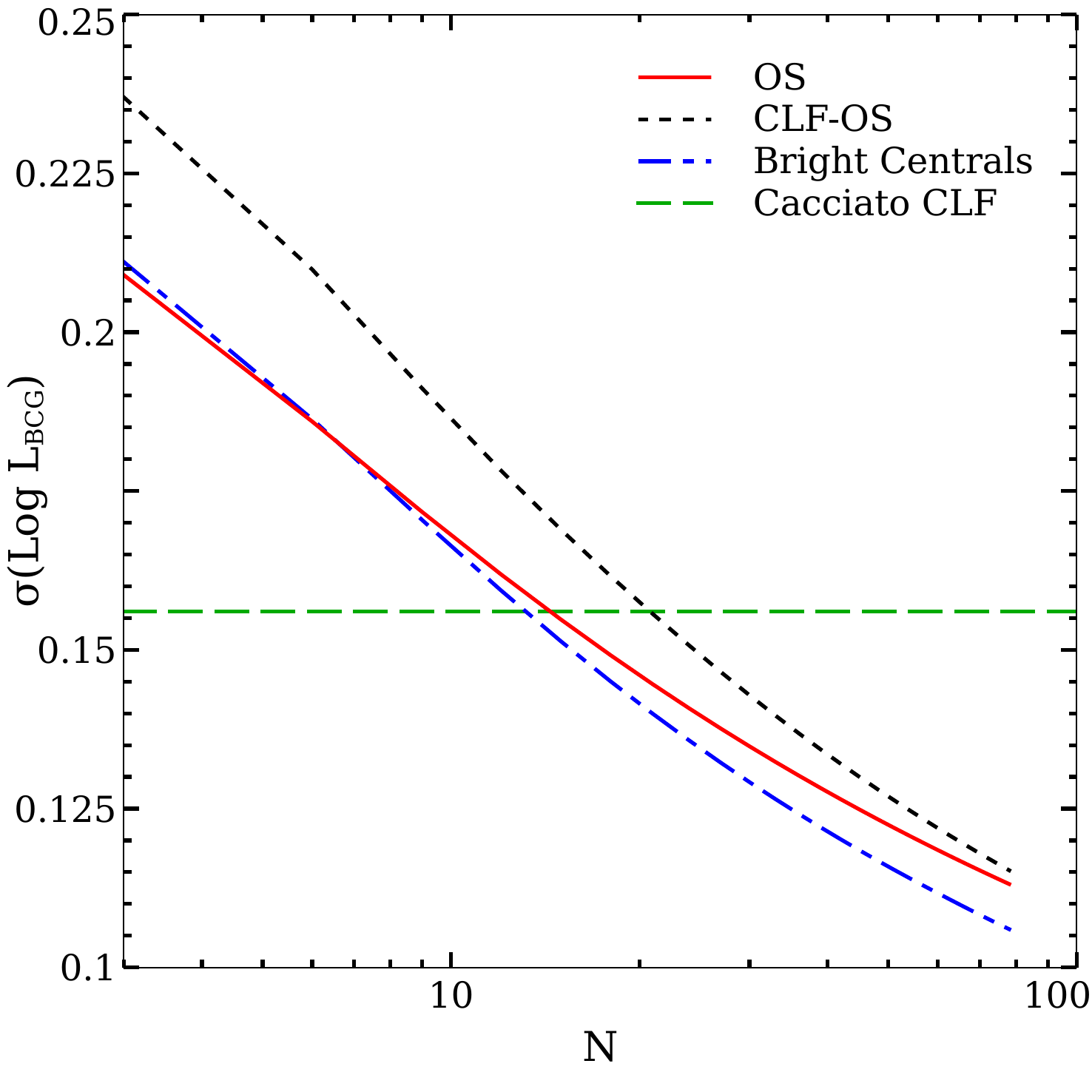}
 \caption{The width of ${\rm lg} \,L_{\rm BCG}$ inferred from the
   distribution $g_1(>L)=1-p(<L)^{N}$ of the luminosity of the
   brightest object in halos containing $N$ galaxies, as a function of 
   $N$ (solid red curve). The dotted black and dot-dashed blue curves show the corresponding mass-averaged predictions of the extended order statistics models of sections~\ref{sec:extreme_value_mass_dependent} and~\ref{sec:centrals_convolved}. The decreasing trend of $\sigma_{{\rm lg}L_{\rm BCG}}$ with $N$ in all these models is similar to the observed one \citep[compare their Figure~13]{Hansen_et_al_2009}, in contrast to the constant width assumed by standard CLF analyses \citep[][dashed green]{Cacciato_2012}.}
\label{fig-sigdex}
\end{figure}

Order statistics also predict how the shape of the BGG luminosity function (rather than just the mean or median) should depend on halo mass. We have already seen in Figure~\ref{fig-lg1l} that the shape of $g_1(L|m)$ is actually not terribly different from a Lognormal, thus providing some justification for the assumption in many CLF analyses that the distribution of central galaxy luminosities is Lognormal. In practice, the predicted distribution of $\ln L_{\rm BGG}$ is slightly narrower at large $\Ns(m)$.  In contrast, most CLF analyses assume the width is independent of $m$. On the other hand, the width is observed to decrease at large $N$ \citep{Hansen_et_al_2009}, and Figure~\ref{fig-sigdex} shows that the order statistics prediction tracks this rather well -- so this may be a case in which the order statistics parametrization is more realistic than the usual CLF analysis.   

In contrast, HOD analyses typically assume that the fraction of halos that host at least one galaxy above some $L$ is given by $[1 + {\rm erf}(y)]/2$ where $y=\log[m/m_0(L)]/\sigma(L)$. The clustering data require that both $m_0$ and $\sigma$ increase as $L$ increases.  Although the appearance of this error function suggests a lognormal distribution, the distribution in $L$ at fixed $m$ is more complicated.  To get a feel for its shape, note that $[1 + {\rm erf}(y)]/2 = 1/2$ when $m=m_0(L)$. Suppose we use $L_{0.5}(m)$ to denote the value of $L$ at which this happens, given an $m$, then this has the appearance of the median BGG luminosity in halos of mass $m$. At large $L$, $m_0(L)$ increases rapidly with $L$, which is consistent with the order statistics expectation that $L_{0.5}$ is a weak function of $m$.  \citep[][e.g., note that $m_0(L)\propto \exp(L/L_0) - 1$, so $L$ grows only logarithmically at large $m$.]{ssm07}

\begin{figure}
 \centering
 \includegraphics[width=0.45\textwidth]{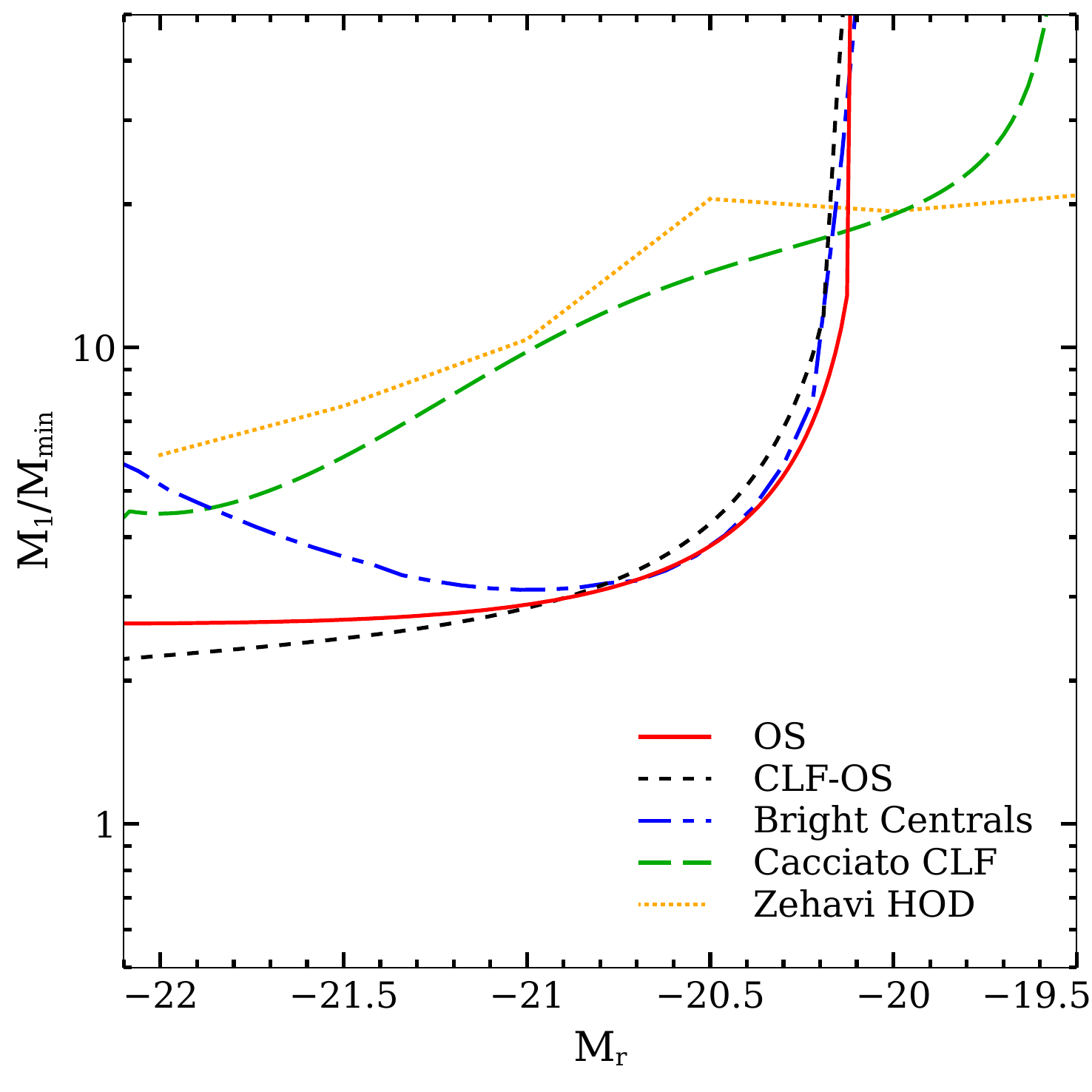}
 \caption{The ratio of the mass scales $M_1(L)$ and $M_{\rm min}(L)$ associated with the satellite and central luminosity functions, respectively (see text for definitions of these scales). The solid red curve shows the order statistics predictions for a universal luminosity function given by the \citet{Blanton_et_al_2003} form and HOD parameters at $M_{r,{\rm max}}=-19.5$ from Z11. The fine-dotted yellow curve shows the Z11 HOD prediction, while the dashed green curve shows the prediction of the standard CLF model of \citet{Cacciato_2012}. For comparison, the dotted black and dot-dashed blue curves show the results of the extended order statistics models described in sections~\ref{sec:extreme_value_mass_dependent} and~\ref{sec:centrals_convolved}. See text for a discussion.}
\label{fig-censatmasses}
\end{figure}

Similarly, if we use $L_{0.84}(m)$ to denote the value of $L$ at which $[1 + {\rm erf}(y)]/2 = 0.84$, then the ratio $L_{0.84}/L_{0.5}$ is a measure of the width of the $L$ distribution at fixed $m$. This happens at $y=1/\sqrt{2} = 0.707$, meaning $\log[m/m_0(L_{0.84})] = \sigma(L_{0.84})/\sqrt{2}$. Since $m=m_0(L_{0.5})$, and $\ln(m_0)\propto L$, we can set $\log[m/m_0(L_{0.84})] \approx (L_{0.5}/L_{0.84})/\ln(10)$, so $(L_{0.84}/L_{0.5}) = [\sqrt{2}/\ln(10)]/\sigma(L_{0.84})$.  
Since $\sigma(L_{0.84})$ increases with $L$, $(L_{0.84}/L_{0.5})$ decreases; the distribution is narrower at large masses.  

\subsubsection{Central and satellite mass scales}
\noindent
The universal order statistics model also predicts that the mass scales determining the central and satellite luminosity distribution must be tightly correlated with one another, at least at large $L$. These mass scales are, respectively, $M_{\rm min}(L)$ defined as the inverse relation of $L_{\rm BGG}(m)$ from before, and $M_1(L)$ which is the scale where $\Ns(>L|m)=1$. To see the correlation between these, consider again the toy model with an exponential luminosity function and $\Ns(m) = (m/\bar M_1)^\alpha$. Simple algebra using \eqns{p0L} and~\eqref{meanNsatL} shows that, in the large $L$ limit where $p(<L)\simeq1$ and $p(>L)\simeq {\rm e}^{-L/L_\ast}$, we have $M_{\rm min}(L)\simeq (\ln2)^{1/\alpha}\,\bar M_1\,{\rm e}^{L/\alpha L_\ast}$ and $M_1(L)\simeq 2^{1/\alpha}\,\bar M_1\,{\rm e}^{L/\alpha L_\ast}$, so that $M_1(L)\simeq 3M_{\rm min}(L)$ if $\alpha\simeq1$. 

Figure~\ref{fig-censatmasses} shows the ratio $M_1(L)/M_{\rm min}(L)$ using order statistics with the \citet{Blanton_et_al_2003} luminosity function and the Z11 HOD parameters at $M_{r,{\rm max}}=-19.5$ (solid red), compared with the HOD prediction for this quantity (fine-dotted yellow) and that in the standard CLF approach using the \citet{Cacciato_2012} model (dashed green, this is computed similarly to the order statistics case, except that the luminosities of centrals do not obey order statistics). For comparison, we also show the predictions of the extended order statistics models of sections~\ref{sec:extreme_value_mass_dependent} and~\ref{sec:centrals_convolved}. 

The universal order statistics model predicts a tight correlation between $M_1(L)$ and $M_{\rm min}(L)$ at large luminosities, as expected from the previous argument, and also predicts that the median BGG luminosity cannot be fainter than a minimum luminosity close to $L_\ast$. In terms of the quantity $m_{1/2}$ introduced earlier, for example, this comes from the fact that the equation $1/2=p(<L)\,{\rm e}^{-\Ns(m)p(>L)}$ has no solution if $L$ is too small. (The extended order statistics models show similar trends.) While the trend at the faint end is qualitatively different from the HOD prediction, at larger $L$ order statistics predict approximately the same slope for $M_1(L)/M_{\rm min}(L)$ as the HOD and standard CLF models, although a substantially different amplitude.

\subsubsection{Spatial clustering}
\begin{figure}
\centering
\includegraphics[width=0.45\textwidth]{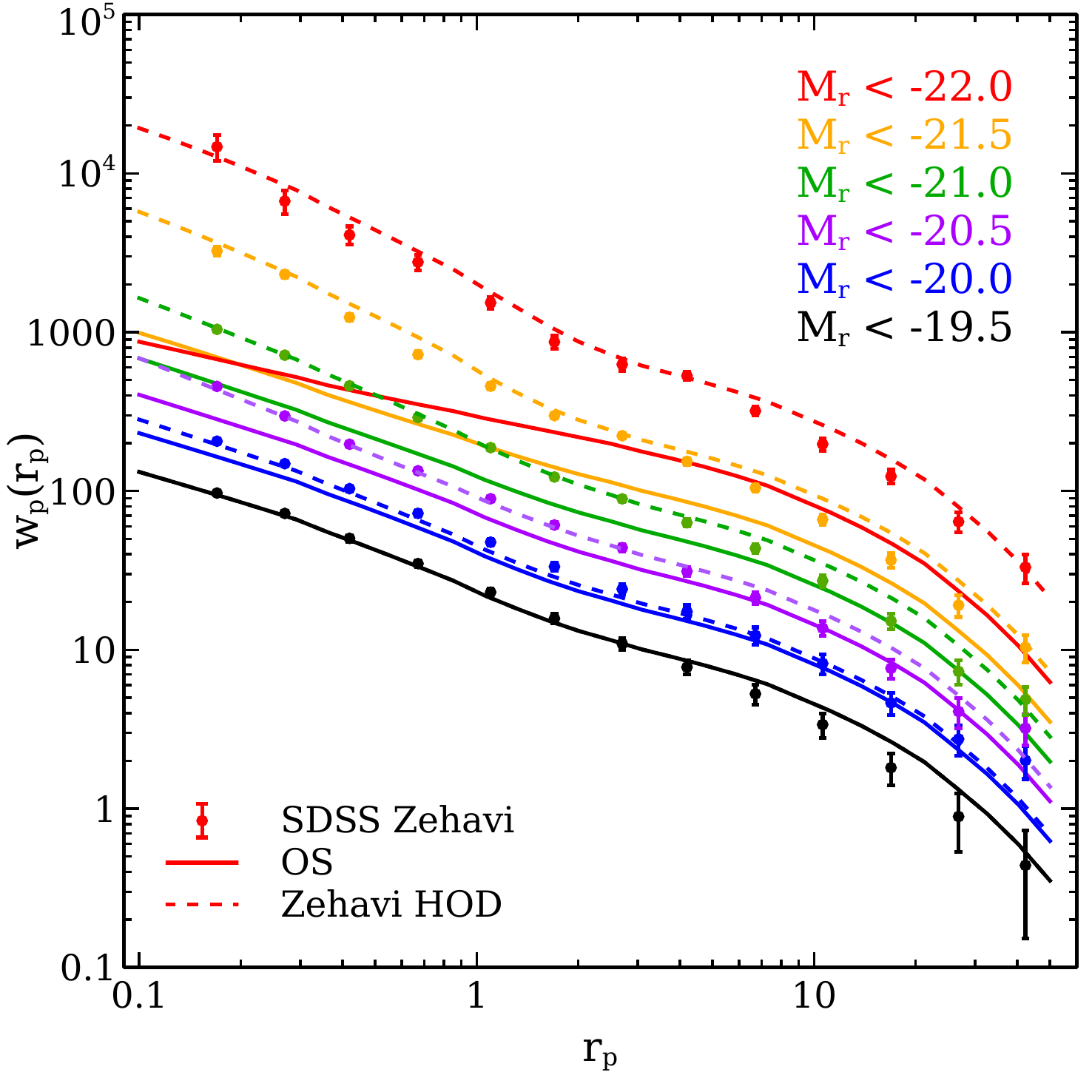}
\caption{Projected correlation function of galaxies. The points with error bars are measurements of $w_{\rm p}(r_{\rm p})$ using SDSS data, taken from Table 8 of \citet[][Z11]{Zehavi_et_al_2011}. The smooth curves show analytical results using the HOD-based model of Z11 with parameters taken from their Table 3 (dotted), and using the order statistics model based on a universal luminosity function as described in the text (solid). Results are shown for galaxy subsamples defined by increasingly bright luminosity thresholds from bottom to top, as labelled. For clarity, all results are staggered by $0.25$dex starting from the sample with $M_r < -20.5$.}
\label{fig:wp_rp_mass_independent}
\end{figure}

\noindent
Figure~\ref{fig:wp_rp_mass_independent} compares the projected correlation function predicted by the universal order statistics model (solid curves) with SDSS measurements reported by Z11 (points with errors) and the HOD model using parameters taken from Table 3 of Z11 (dotted curves). We have used the luminosity function of \cite{Blanton_et_al_2003} to calculate $p(L)$ for the order statistics model. In all calculations of $w_{\rm p}(r_{\rm p})$, we set the upper limit of the integral in \eqn{eq:wp_rp} to $r_{\pi}^{\rm max} = 60\hMpc$, consistent with Z11.
Clearly, at higher luminosities this model predicts systematically less clustering than is observed. It is easy to see why this is so at large scales ($k\to0$), where the shape of the satellite profile inside haloes plays no role ($u(k|m)\to1)$. This means the integral in the 2-halo term \eqref{eq:P2h} involves the mean \emph{total} number of galaxies $\fcen(>L|m)+\fcen(m)\Ns(>L|m)=\avg{N|>L,m}$ (equation~\ref{meanNL-censat}). Since the luminosity dependence of the latter is simply an overall factor of $p(>L)$ (equation~\ref{meanNL}), this cancels with the corresponding luminosity dependence of $\ngal(>L)$ (equation~\ref{eq:avg_ng}), leaving \emph{no} luminosity dependence of large scale clustering.

Despite the promising trends discussed earlier, therefore, this model based on a universal luminosity function fails rather dramatically at reproducing the observed luminosity dependence of clustering. This is consistent with the conclusions of \cite{Paranjape_Sheth_2012} based on marked correlation statistics of SDSS groups. We explore possible improvements to this model in the sections that follow. 

\section{Order statistics with a Conditional Luminosity Function}
\label{sec:extreme_value_mass_dependent}
\noindent
The simplest extension to the model based on a universal luminosity function is to allow the luminosity function that drives order statistics in a group to depend on the mass of the group's parent dark matter halo. In other words, the distribution $p(L)$ from earlier must be replaced with $p(L|m)$, with no other change in any of the expressions for counts and correlations. The Conditional Luminosity Function (CLF) approach \citep{Yang_et_al_2003, Yang_et_al_2008, Surhud_2012, Cacciato_2012, V_Bosch_et_al_2013}, which we briefly describe next, provides a natural language in which to present the results. This basic theme has also been considered previously by other authors \citep{Vale_Ostriker_2006,Vale_Ostriker_2008}.

\begin{figure}
\centering
\includegraphics[width=0.45\textwidth]{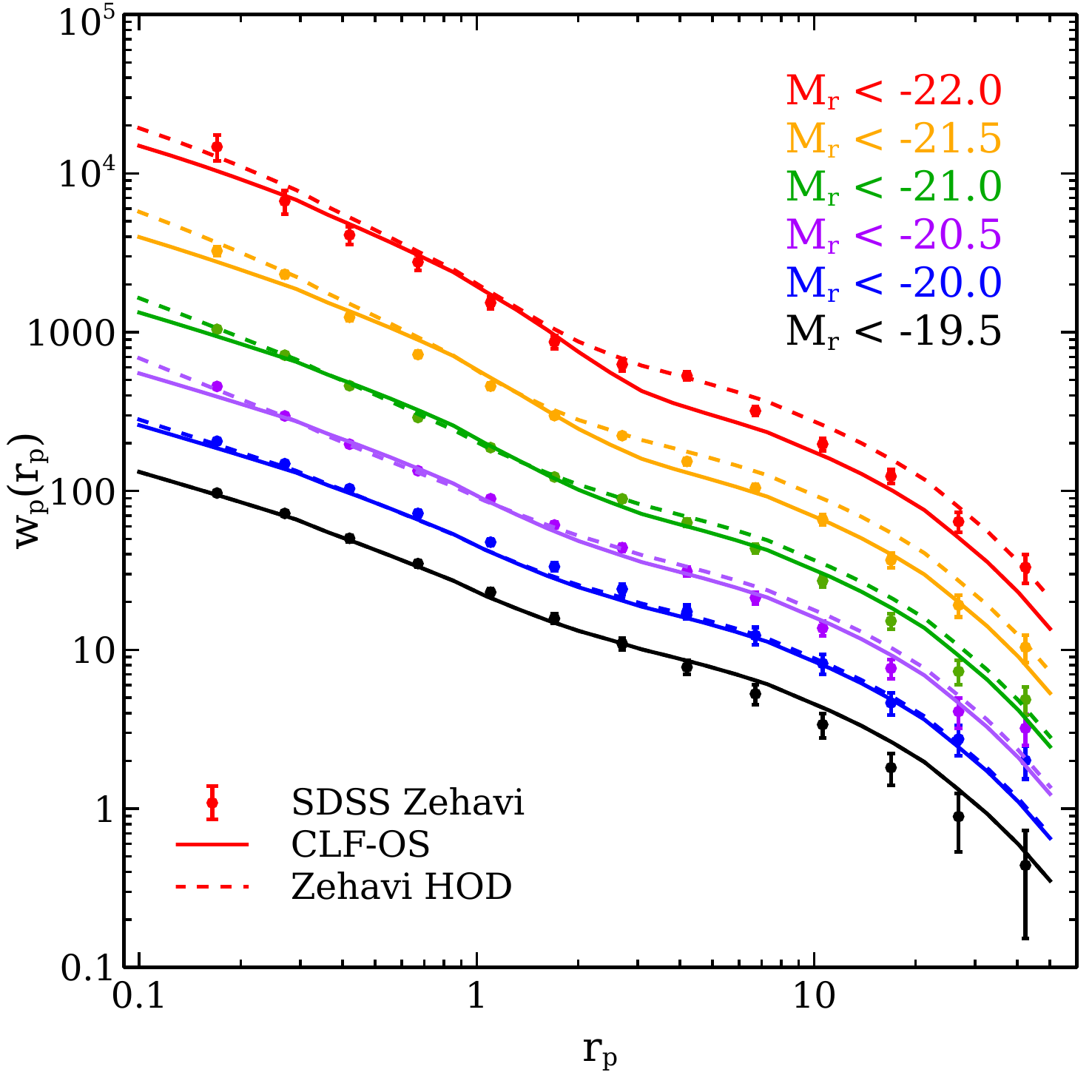}
\caption{Same as Figure~\ref{fig:wp_rp_mass_independent}, except that the solid curves now show the results of the CLF-based order statistics model as discussed in the text. This model is clearly in much better agreement with the data than the one based on a universal luminosity function $p(L)$. }
\label{fig:wp_rp_mass_dependent}
\end{figure}

\begin{table}
\begin{center}
\begin{tabular}{cccc}
\hline
Parameter & CLF-OS & Bright Centrals & Equation \\[.5 ex]
\hline \hline
$M_r^{\rm max}$ & -19.5 & -19.5 & \\[.5 ex]
\hline
$\beta$ & 1.2 & 1.25 & \eqref{eq:phi_L_m} \\[.5 ex]
\hline
$M_r^0$ & -20.44 & -20.44 & \eqref{eq:expression_L_star} \\[.5 ex]
\hline
$\log m_1 $ & 9.81 & 9.81 & \eqref{eq:expression_L_star} \\[.5 ex]
\hline
$\gamma_1$ & 2.75 & 2.75 & \eqref{eq:expression_L_star} \\[.5 ex]
\hline
$\gamma_2$ & 0.0672 & 0.06 & \eqref{eq:expression_L_star} \\[.5 ex]
\hline \hline
\end{tabular}
\end{center}
\caption{Parameter values used in this work for the CLF-based order statistics model (column 2) and for the order statistics model with central galaxies brightened (column 3). Note that $M_r^0$ is the absolute magnitude corresponding to $L_0$. See text for details.}
\label{table:best_fit_exHOD_both_models}
\end{table} 

The halo mass-dependent luminosity distribution $p(L|m)$ can be written using the CLF $\phi(L|m)$ as
\be
p(L|m) = \phi(L|m) / \phi(>L_{\rm min}|m)\,,
\label{eq:p(L|m)-def}
\ee
where the mass dependence of the CLF must be calibrated against data, the difference with the usual CLF approach being that we are not explicitly distinguishing between centrals and satellites. The CLF is defined such that the overall luminosity function of the galaxies is given by
\be
\phi(L) = \int \der m\,n(m)\, \phi(L|m)\,. 
\label{eq:CLF} 
\ee
Following \citet{Yang_et_al_2008}, we choose
\begin{align}
\phi(L|m) &= \frac{\phi_\ast(m)}{L_\ast(m)}\left(\frac{L}{L_\ast(m)}\right)^\alpha \exp \left[ -\left( \frac{L}{L_\ast(m)} \right)^\beta \right] \,\, , \label{eq:phi_L_m} 
\end{align}
where
\begin{align} 
L_\ast(m)=L_0\frac{(m/m_1)^{\gamma_1}}{\left( 1+ m/m_1\right)^{\gamma_1-\gamma_2}} \,, 
\label{eq:expression_L_star}
\end{align} 
and the mass dependence of $\phi_\ast$ is fixed by demanding that, for the lowest luminosity threshold $\Lmin$, the cumulative average number of galaxies $\avg{\Ngal|>\Lmin,m}$ matches the standard HOD expression $f_{\rm cen}(m)[1+\Ns(m)]$ appropriate for this threshold. Demanding that the faint end slope of the luminosity function match exactly that of the Schechter function parametrisation of \cite{Blanton_et_al_2003} allows us to set $\alpha\equiv\alpha_{\rm Schechter}=-1.05$.  Therefore we have five free parameters $L_0$, $\beta$, $m_1$, $\gamma_1$ and $\gamma_2$ left  in the model which are independent of halo mass and galaxy luminosity. The universal luminosity function model described in the previous section is recovered by simply setting $\beta = 1$ and $\gamma_1 = \gamma_2 = 0$ in \eqns{eq:phi_L_m} and \eqref{eq:expression_L_star}.

We have determined the values of these parameters by trial and error so as to match the computed projected correlation function $w_{\rm p}(r_{\rm p})$, the overall galaxy luminosity function $\phi(L)$, the luminosity distribution functions of the brightest group galaxy (BGG) and the second brightest group galaxy (SBGG) with the observed data; the values are reported in Table \ref{table:best_fit_exHOD_both_models}.

\begin{figure}
\centering
\includegraphics[width=0.45\textwidth]{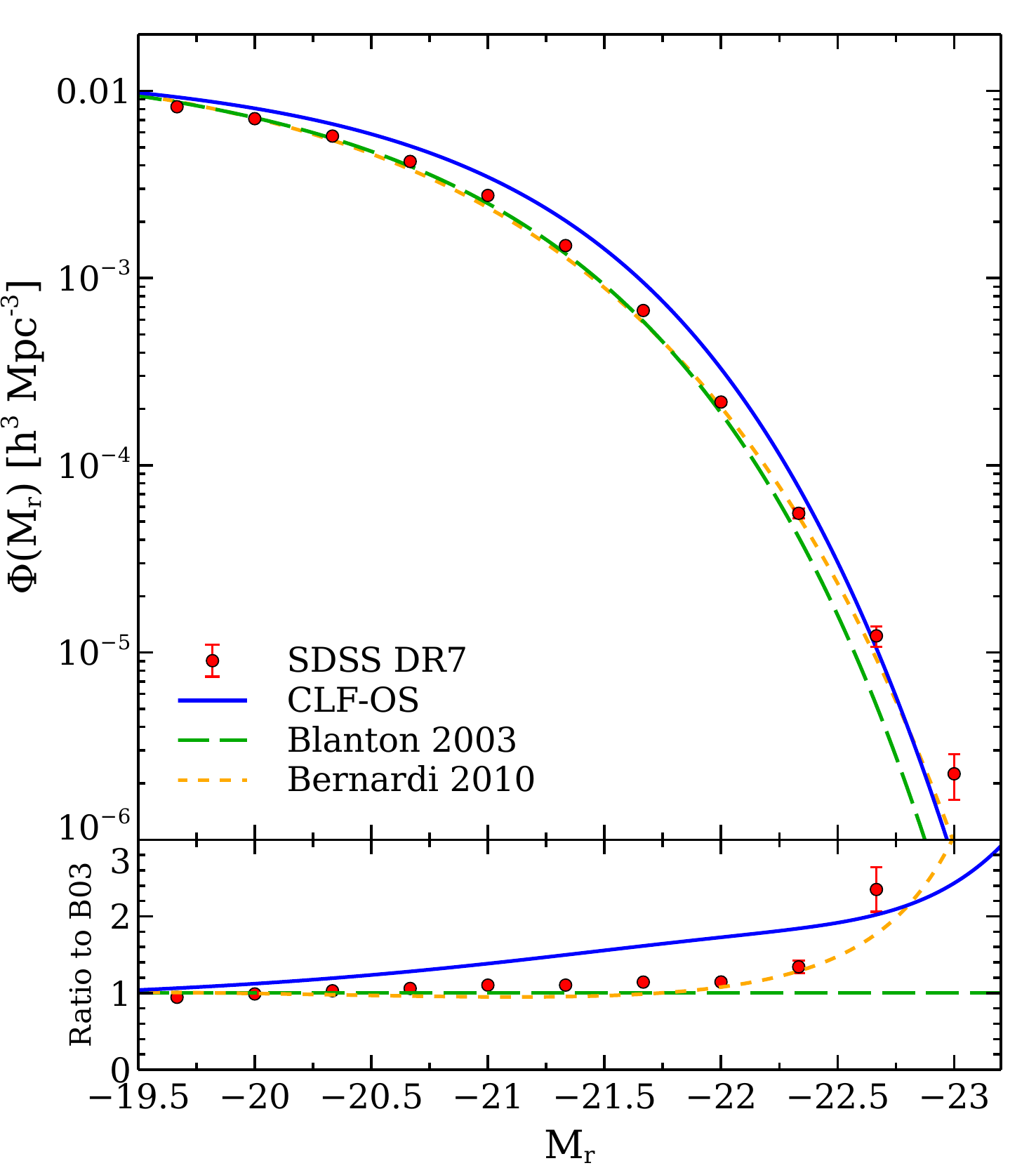}
\caption{Luminosity function of all galaxies brighter than $M_{r,{\rm max}}=-19.5$. \emph{(Upper panel):} The red circles represent measurements from the SDSS DR7 sample described in the text, with error bars computed using 150 bootstrap samples. The solid blue curve represents the luminosity function computed in our CLF-based model using \eqn{eq:CLF}, \eqref{eq:phi_L_m} and \eqref{eq:expression_L_star}. The dashed green and the dotted orange curves show the Schechter function fit of \citet{Blanton_et_al_2003} and modified Schechter fit of \citet{Bernardi_et_al_2010}, respectively. \emph{(Lower panel):} Ratios of the various luminosity functions to that of \citet{Blanton_et_al_2003}.}
\label{fig:phi_M_exHOD_mass_dependent}
\end{figure}

Figure~\ref{fig:wp_rp_mass_dependent} shows the projected correlation function of galaxies. The points with errors and the dotted curves are repeated from Figure~\ref{fig:wp_rp_mass_independent}.
The solid curves show the correlation function computed in our CLF-based order statistics model using appropriately modified versions of \eqn{eq:P2h}, \eqref{eq:xi_1h(r)} and \eqref{eq:wp_rp}. Figure~\ref{fig:phi_M_exHOD_mass_dependent} shows the all-galaxy luminosity function. The circles represent the luminosity function measured from the SDSS subsample described earlier and the error bars are calculated using 150 bootstrap samples.

\begin{figure*}
\centering
\includegraphics[width=0.45\textwidth]{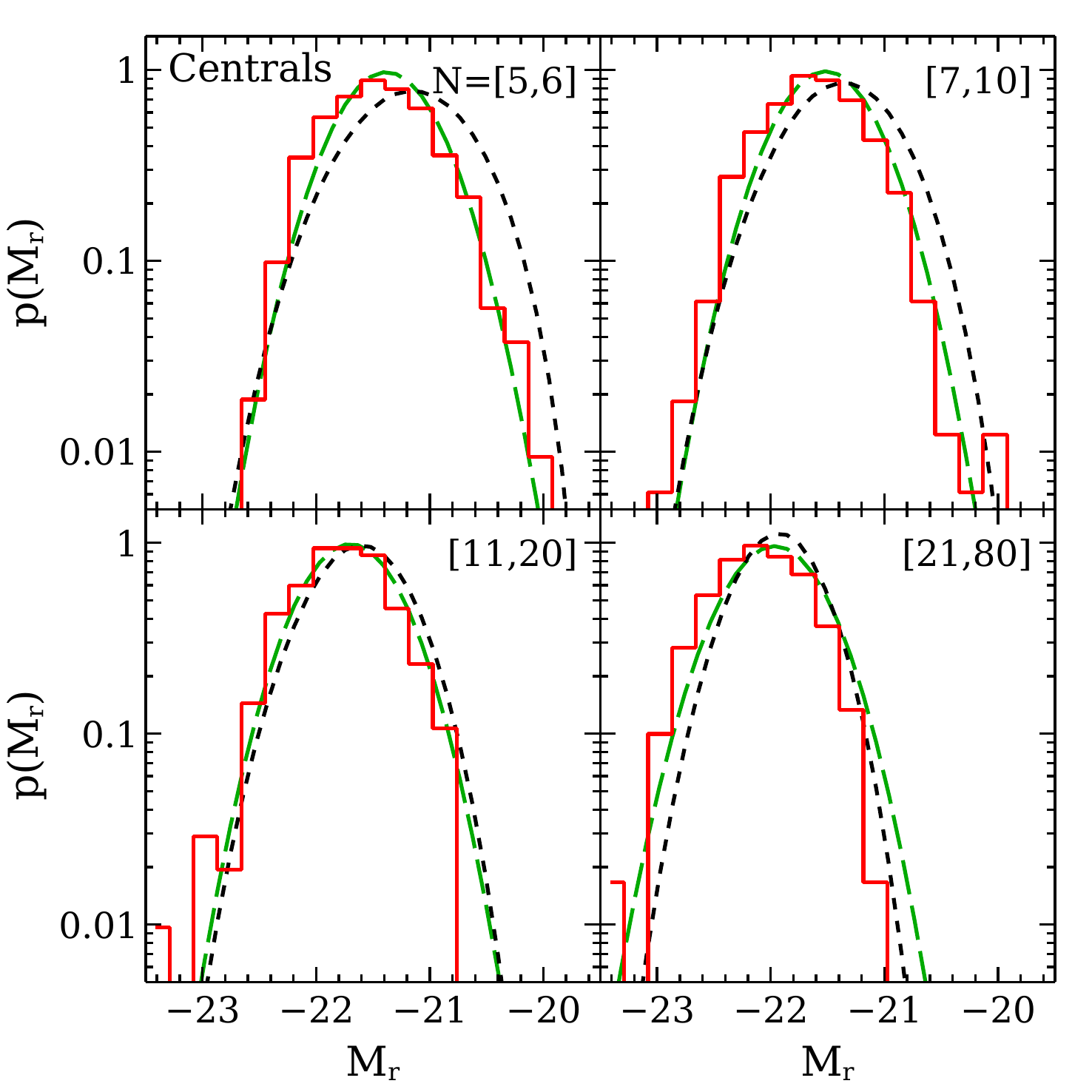}
\includegraphics[width=0.45\textwidth]{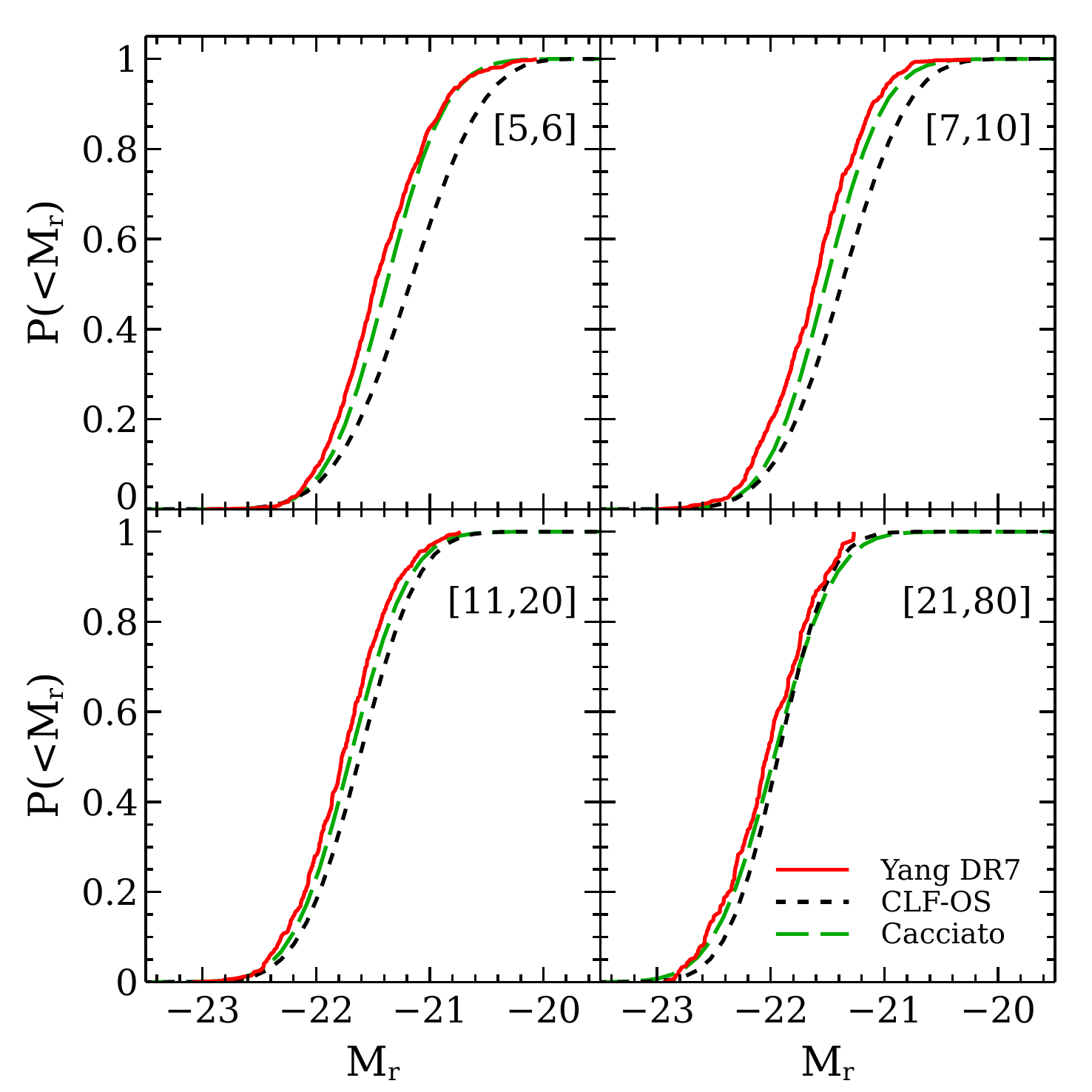}
\caption{Luminosity distribution of the brightest group galaxies in four different richness bins (differential distributions in the \emph{left panel} and the corresponding cumulative distributions in the \emph{right panel}). The solid red histograms show measurements from the group catalogue of \citet{Yang_et_al_2007} and the black dotted curves represent the luminosity distribution calculated using our CLF-based order statistics model using \eqns{eq:differential_g_N_n}, \eqref{eq:p(L|m)-def} and \eqref{eq:brightness_distribution_binned_version} with $n=1$. We see that this model systematically produces fainter central galaxies than observed. For comparison, we also show the luminosity distribution of the centrals galaxies calculated from the CLF approach of \citet[][dashed green curves]{Cacciato_2012}.} \label{fig:BGG_exHOD_all_bins}
\end{figure*}

\begin{figure*}
\centering
\includegraphics[width=0.45\textwidth]{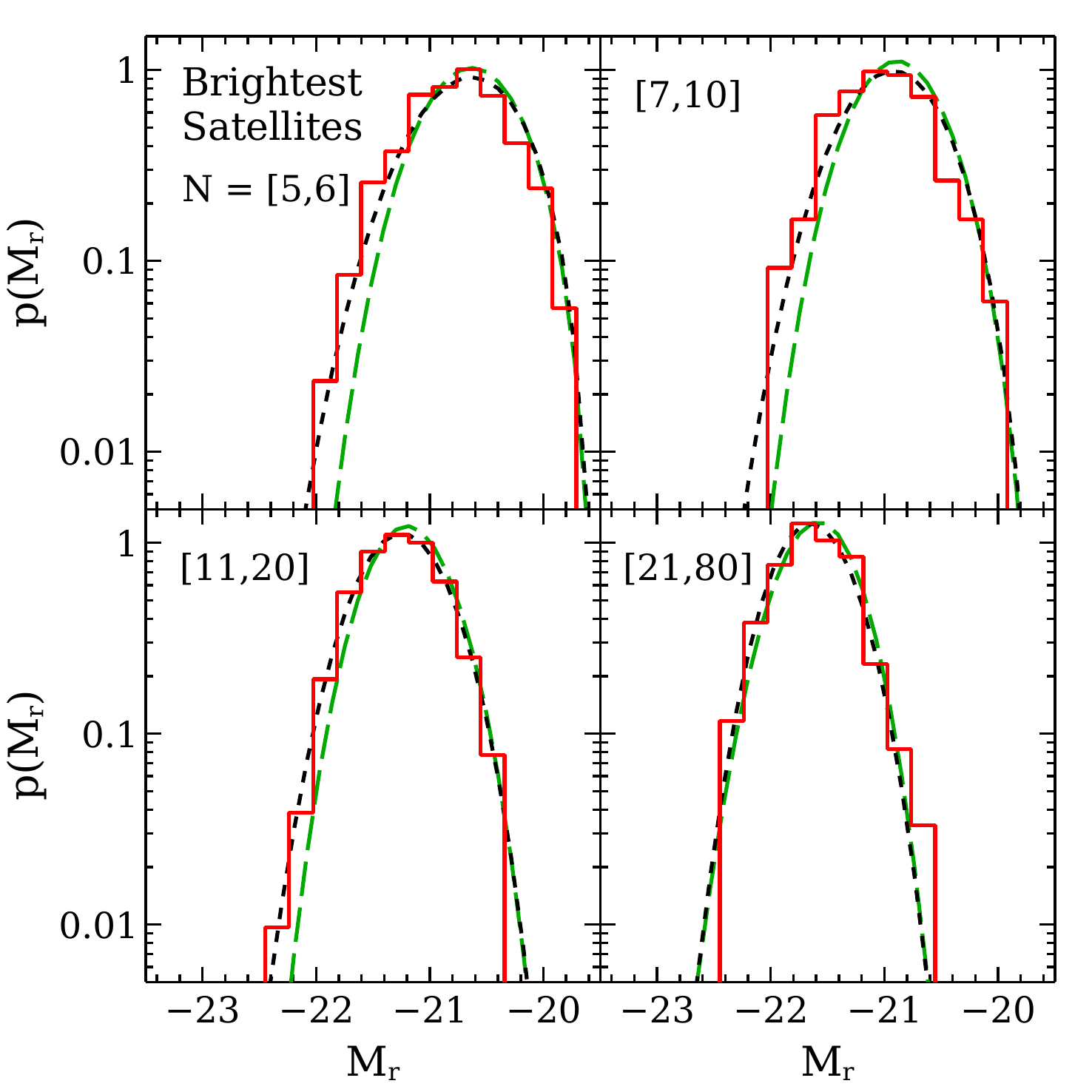}
\includegraphics[width=0.45\textwidth]{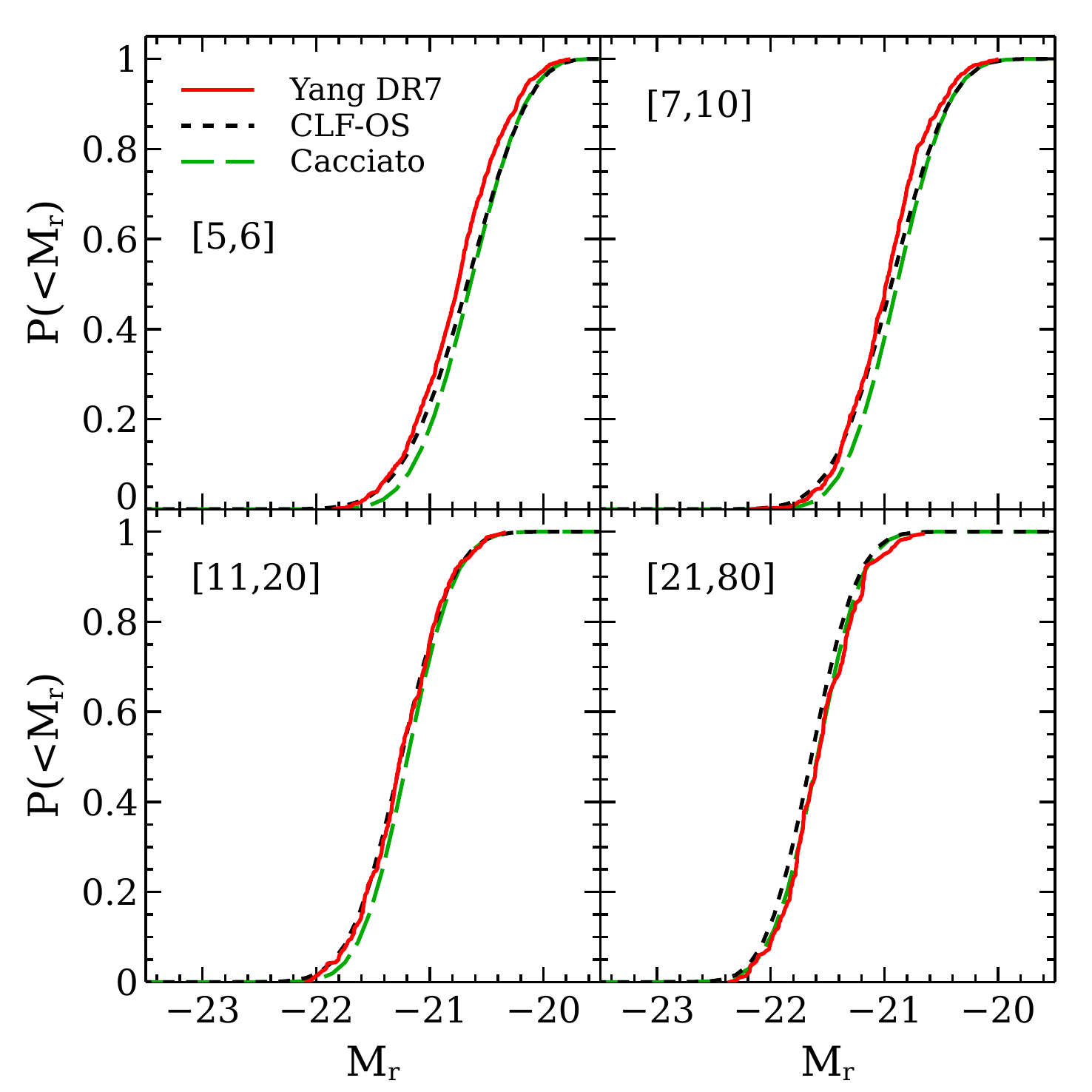}
\caption{Similar to Figure~\ref{fig:BGG_exHOD_all_bins}, now showing luminosity distributions of the second brightest galaxies. The black dotted curves represent the SBGG luminosity distribution calculated using our CLF-based order statistics model using \eqns{eq:differential_g_N_n}, \eqref{eq:p(L|m)-def} and \eqref{eq:brightness_distribution_binned_version} with $n=2$. We see that this model produces the SBGG distribution in good agreement to the data. For comparison, we also show the luminosity distribution of the brightest satellites calculated from the CLF approach of \citet[][dashed green curves]{Cacciato_2012}.} \label{fig:SBGG_exHOD_all_bins}
\end{figure*}

Although the projected correlation function and the all-galaxy luminosity function in this model are reasonably consistent with the observed data, the model systematically predicts fainter central galaxies than are observed, as shown in Figure~\ref{fig:BGG_exHOD_all_bins}. In both the panels of this figure, the dotted black curves denote our model based on halo mass dependent extreme value statistics and the solid red histograms represent measurements from the Yang et al. group catalog. The luminosity distribution of the \emph{second} brightest group galaxies, however, agrees well with corresponding measurements of brightest satellite luminosities from the Yang et al. group catalog as shown in Figure~\ref{fig:SBGG_exHOD_all_bins}. 

To calculate the luminosity distributions in this case, we replace $p(L)$ and its cumulative forms in \eqn{eq:differential_g_N_n} with $p(L|m)$ and the corresponding cumulative forms. The luminosity distributions of the BGG and SBGG correspond to $n = 1$ and $n = 2$ of that equation, respectively. But the quantities we plot in Figures~\ref{fig:BGG_exHOD_all_bins} and~\ref{fig:SBGG_exHOD_all_bins} are independent of halo mass and they are computed in different bins of group richness instead of at a fixed richness value. This is obtained by averaging over halo mass and binning in richness:
\begin{align}
&g_n(L| N_1<N<N_2) \notag\\
&\ph{g_n}= \frac{\sum_{N = N_1}^{N_2} \int \der m\,n(m)\, g_n(L|m,N)\,p(N|m)}{\sum_{N=N_1}^{N_2} \int \der m\,n(m)\, p(N|m)} \,. 
\label{eq:brightness_distribution_binned_version}
\end{align}
Here $p(N|m)$ denotes the probability to find $N$ galaxies (i.e. $N-1$ satellites) in a halo of mass $m$; as before, we choose this to be a Poisson distribution with mean satellite number $\Ns$. Note that the expression of \eqn{eq:brightness_distribution_binned_version} is true not only for $g_n$ but for any probability density function which is defined at fixed halo mass and group richness. 

For comparison, the luminosity distributions of the centrals and brightest satellites calculated from the CLF approach of \citet{Cacciato_2012} is shown by the dashed green curves in Figures~\ref{fig:BGG_exHOD_all_bins} and \ref{fig:SBGG_exHOD_all_bins}, respectively. In this approach, one splits the conditional luminosity function into its central and satellite part so that, $\phi(L|m) = \phi_{\rm cen}(L|m) + \phi_{\rm sat}(L|m)$. Therefore the luminosity distribution of the centrals is $p_{\rm cen}(L|m) = \phi_{\rm cen}(L|m)/\phi_{\rm cen}(>\Lmin|m)$. The brightest satellite of a group with richness $N$ in this formalism can be mapped to the brightest galaxy of a group of $N-1$ galaxies, all of which have their luminosities drawn from the luminosity distribution $p_{\rm sat}(L|m) = \phi_{\rm sat}(L|m)/\phi_{\rm sat}(>\Lmin |m)$. Therefore we can use \eqn{eq:differential_g_N_n} with $n=1$ to calculate the luminosity distribution of the brightest satellite in this case.

In summary, we see that introducing a halo mass dependence in the luminosity function underlying the order statistics hypothesis, $p(L)\to p(L|m)$, does in fact lead to a dramatically improved comparison with a number of observables. Importantly, though, this model produces systematically fainter central galaxies than are observed, consistent with some additional, unmodelled physics that make the central galaxy special. In the next section we statistically model this unknown physics using an \emph{ad hoc} brightening of the central galaxy, over and above the order statistics prediction.

\section{Order Statistics with Centrals Brightened} \label{sec:centrals_convolved}
\noindent
The fact that the halo mass dependent order statistics model of the previous section predicts fainter central galxies than are observed is consistent with the results of \citet{Shen_et_al_2014}, who showed that an order statistics model using a luminosity function that depends on group richness was unable to match the BGG luminosity distribution. \citet{Shen_et_al_2014} therefore proposed a model in which central luminosities were brightened by an amount depending on the observed magnitude gap of the group. 

We will pursue a somewhat different approach here. Instead of using the observed statistics of the magnitude gap, we will \emph{forward model} the central luminosity distribution by convolving the BGG luminosity distribution predicted by order statistics with an appropriate kernel, while keeping satellite luminosities unchanged. An advantage of this formulation is that the physics that makes the centrals special is now statistically described by the brightening kernel (equation~\ref{eq:kernel} below). This provides a useful language for comparing Halo Model predictions with more physically motivated semi-analytical models of galaxy evolution \citep[see][for a review]{Somerville_Dave_2015}. As before, we demand agreement between the model predictions and observations for the all-galaxy luminosity function, the 2-point correlation function and the luminosity distributions of the brightest and second brightest group galaxies. The magnitude gap distribution -- discussed in the next section -- is then a \emph{prediction} of this model. 

The order statistics prediction for the BGG luminosity distribution is given by \eqn{eq:differential_g_N_n} with $n=1$. Now to brighten the centrals, we modify this distribution using
\begin{align}
p_{\rm cen}(M|m,N) = \int_{-\infty}^{M_{\rm max}} \der M^{\prime} G(M,M^{\prime})g_1(M^{\prime}|m,N), 
\label{eq:convolution_g1}
\end{align}
where $M$ denotes the absolute magnitude of the central galaxies and the luminosity distribution as a function of absolute magnitude satisfies (with some abuse of notation) $p(M)=L(M)p(L(M))\ln(10)/2.5$ and $p(>M)=p(<L)$. The kernel $G$ is modelled as,
\begin{align}
G(M,M^{\prime}) = 
\frac{{\rm e}^{- \left( M + \mu - M^{\prime} \right)^2/2 \sigma^2}/\sqrt{2\pi}\sigma}{ \frac12\left\{1 +  \erf{\frac{M_{\rm max}+\mu-M^{\prime}}{\sqrt{2}\sigma}}\right\}} \,,
\label{eq:kernel}
\end{align}
where the denominator ensures the normalisation $\int^{M_{\rm max}}_{-\infty}\der M\, p_{\rm cen}(M|m,N) = 1$, and $M_{\rm max}$ is the absolute magnitude corresponding to the luminosity threshold $L_{\rm min}$. 

In the limit $\sigma\to0$, the convolution in \eqn{eq:convolution_g1} would amount to a constant shift $M^\prime\to M=M^\prime-\mu$, which is a brightening when $\mu > 0$. We find that the following prescription for $\mu$ and $\sigma$ gives a reasonable match to the observables. We perform the convolution only for $\log m >10.2$, using $\sigma=0.01$ and
\begin{align}
\mu &= 0.09\times\erfc{\frac{\log m - 14.0}{0.8}}\notag\\
&\ph{0.09\times[\log]}
\times\left[1 + \erf{\frac{\log m - 12.0}{0.8}}\right]
\label{eq:expression_mu} 
\end{align}
and further use the (now slightly modified) parameter values for the conditional luminosity function from the third column of Table~\ref{table:best_fit_exHOD_both_models}. Thus, in this model we only brighten BGGs that live in haloes of mass $m\sim10^{13}h^{-1}\Msun$. This will have interesting consequences for the distribution of magnitude gap; we discuss these, as well as the robustness of our choice of brightening kernel, in the next section.

\begin{figure*}
\centering
\includegraphics[width=0.45\textwidth]{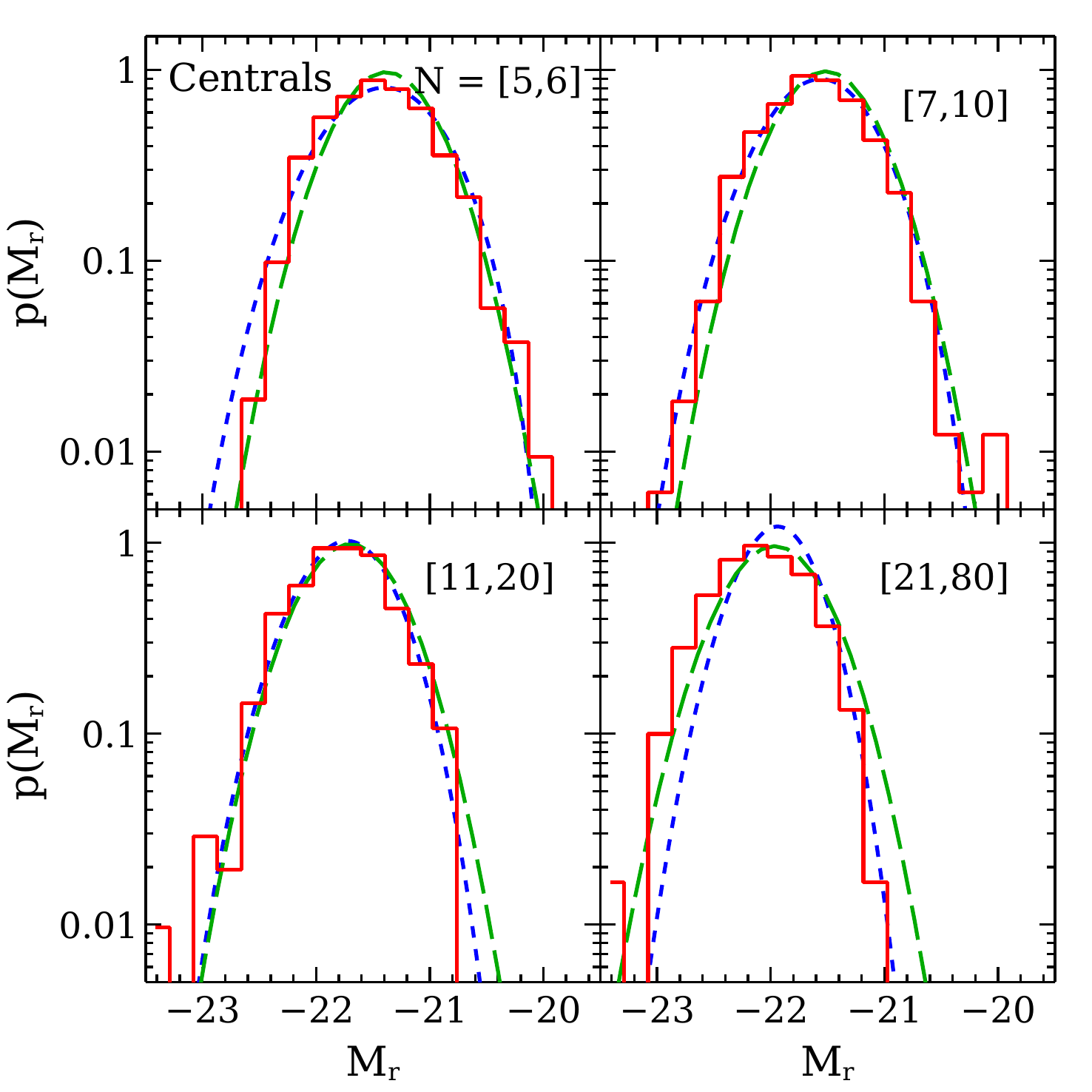}
\includegraphics[width=0.45\textwidth]{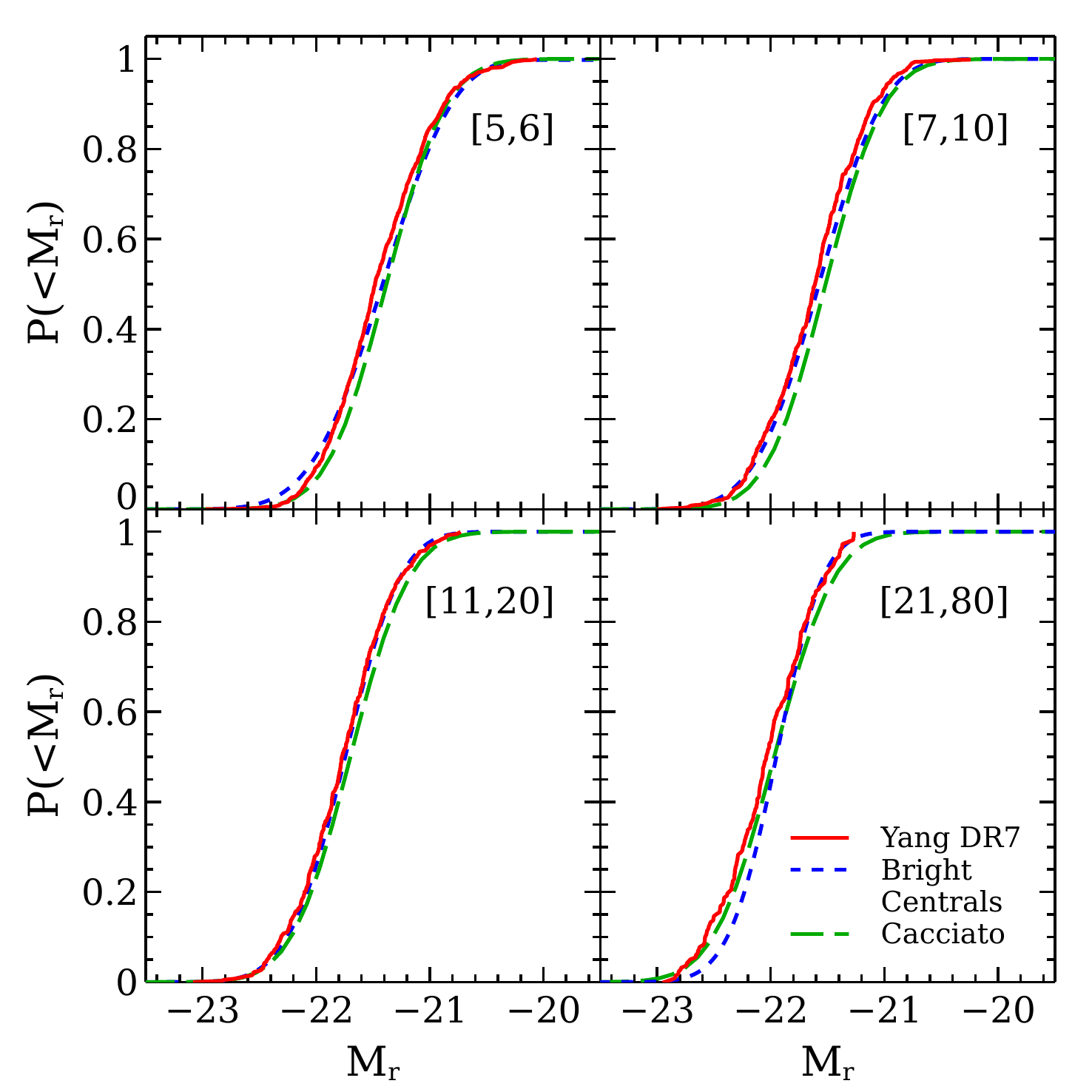}
\caption{Same as Figure~\ref{fig:BGG_exHOD_all_bins}, except that we now show the central luminosity distributions computed using \eqn{eq:convolution_g1} as the dotted blue lines. We see that this new model, in which the BGG luminosities are additionally brightened using the kernel \eqref{eq:kernel}, is now consistent with the data for all values of group richness.}  \label{fig:BGG_centrals_convolved}
\end{figure*}

In Figure~\ref{fig:BGG_centrals_convolved} we show the luminosity distribution of the centrals computed using \eqn{eq:convolution_g1}, compared with the Yang et al. group catalog measurements and CLF predictions of \citet{Cacciato_2012} (repeated from Figure~\ref{fig:BGG_exHOD_all_bins}). We see that the predicted luminosity distribution of centrals now agrees very well with the data at all values of group richness. 

\begin{figure*}
\centering
\includegraphics[width=0.45\textwidth]{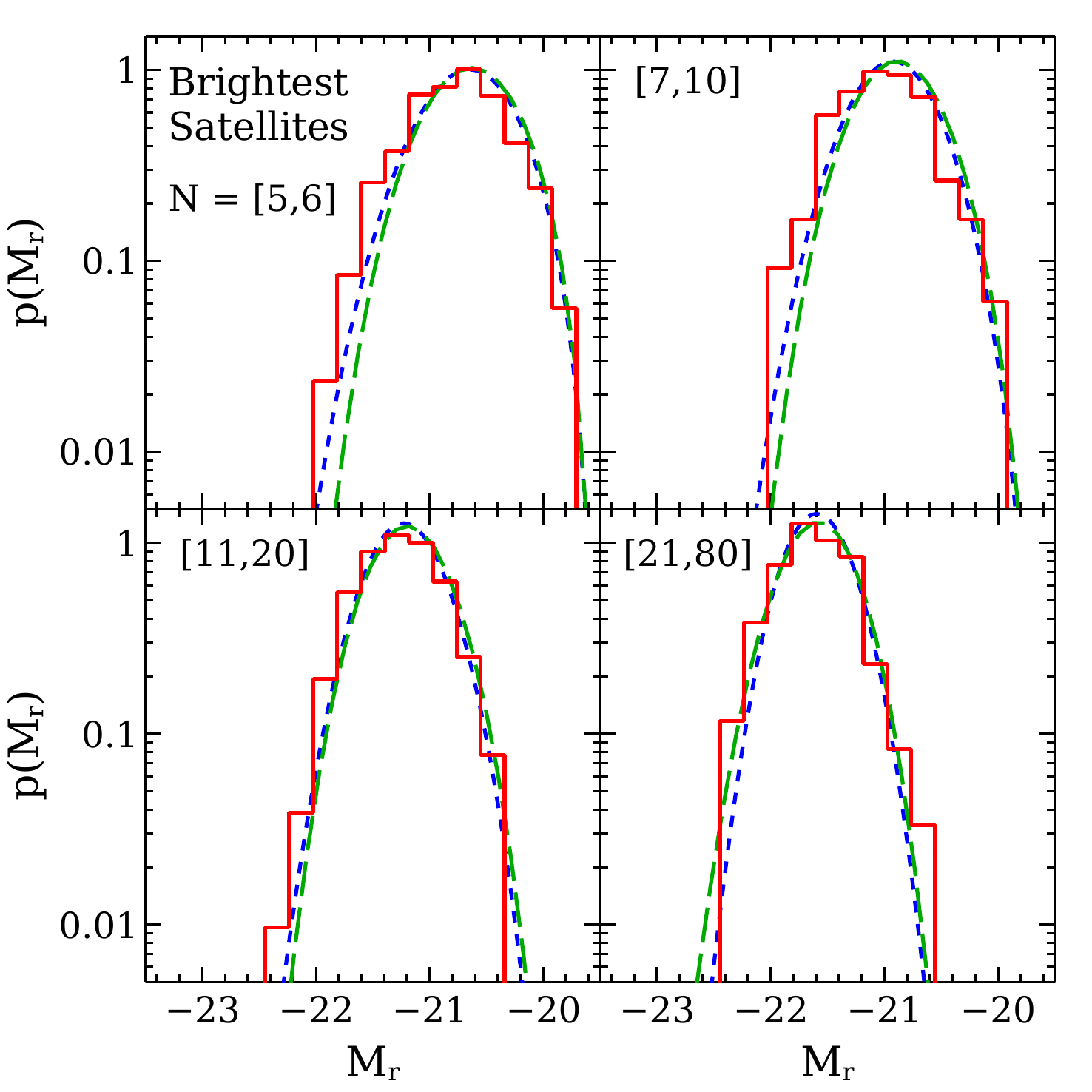}
\centering
\includegraphics[width=0.45\textwidth]{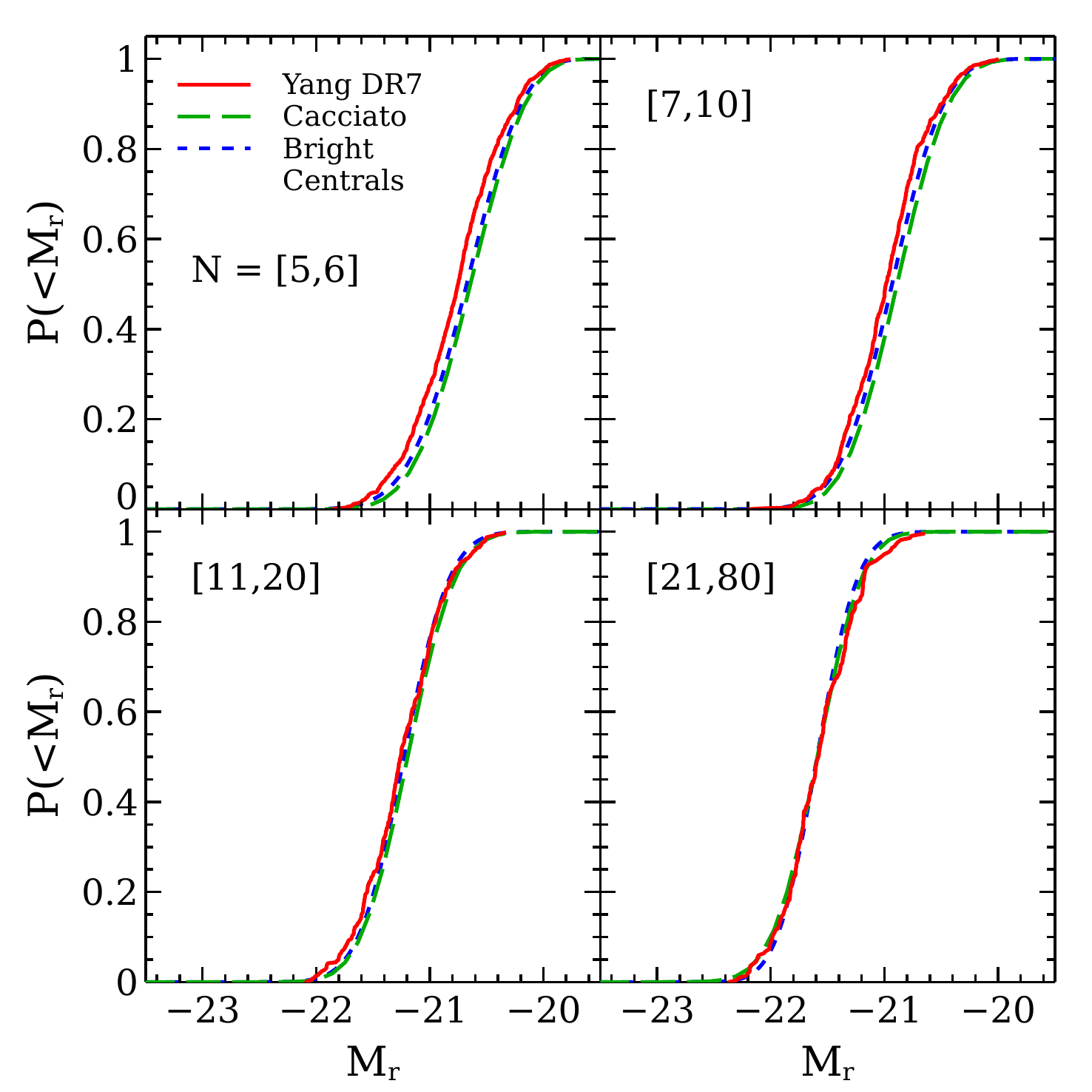}
\caption{Same as Figure~\ref{fig:SBGG_exHOD_all_bins}, except that we now show the brightest satellite luminosity distributions computed using \eqref{eq:p_sat} in \eqn{eq:differential_g_N_n} with $n=1$ as the dotted blue lines. We see that the BSG luminosities in this new model are still consistent with the data.}  \label{fig:BSG_centrals_convolved}
\end{figure*}

Figure~\ref{fig:BSG_centrals_convolved} shows the luminosity distribution of the brightest satellites. In our new model we keep the satellite luminosities unchanged, with a distribution given by \eqn{eq:p_sat}. So the brightest satellite in group of $N$ galaxies can be thought of as the brightest galaxy in a group of $N-1$ galaxies, all of which have their luminosities drawn from $p_{\rm sat}(L|m,N)$ (equation \ref{eq:p_sat}), which we therefore use in \eqn{eq:differential_g_N_n} with $n=1$. The measurements from the Yang et al. catalogue and the standard CLF results using the \citet{Cacciato_2012} model are repeated from Figure~\ref{fig:SBGG_exHOD_all_bins}. We see that the computed luminosity distribution of the brightest satellites is also consistent with the data.

Now we discuss how to calculate the projected correlation function and the all-galaxy luminosity function in this new model. Since after brightening the centrals we are no longer in the order statistics domain, we have to modify our definition of HOD. Now we will define our HOD in the standard way (equation \ref{eq:Ng_avg_standard_HOD}) with the two quantities $\fcen^{\rm new}(>L|m)$ and $\Ns^{\rm new}(>L|m)$. We can find the expression for $\fcen^{\rm new}(>L|m)$ in the following way, 
\begin{align}
\fcennew (>L|m) &= \fcen(m)p_{\rm cen}(>L|m) \,, 
\label{eq:fcen_new}
\end{align} 
where we recall that $\fcen(m)=\fcen(>\Lmin |m)$. Here $p_{\rm cen}(>L|m)$ denotes the probability to find a central galaxy brighter than $L$ in a halo of mass $m$, and is the weighted average of $p_{\rm cen}(>L|m,N)$ defined in \eqn{eq:convolution_g1}. In our model, we have brightened the centrals but have kept the luminosity of the satellites intact. Hence, following the two definitions of HOD, one in  \eqn{meanNL-censat}  and the other in \eqn{eq:Ng_avg_standard_HOD}, we can write, 
\begin{align}
\Nsnew (>L|m) &= \frac{\fcen(m)}{\fcennew(>L|m)}\Ns(>L|m) \label{eq:Ns_new} \,\, ,
\end{align} 
Having defined HOD in this way, we can use \eqn{eq:P2h}, \eqref{eq:xi_1h(r)} and \eqref{eq:wp_rp} to calculate the projected correlation function in this framework, after modifying those equations as is done in the standard HOD framework (see the discussion below equation \ref{eq:xi_1h(r)}).

\begin{figure}
\centering
\includegraphics[width=0.45\textwidth]{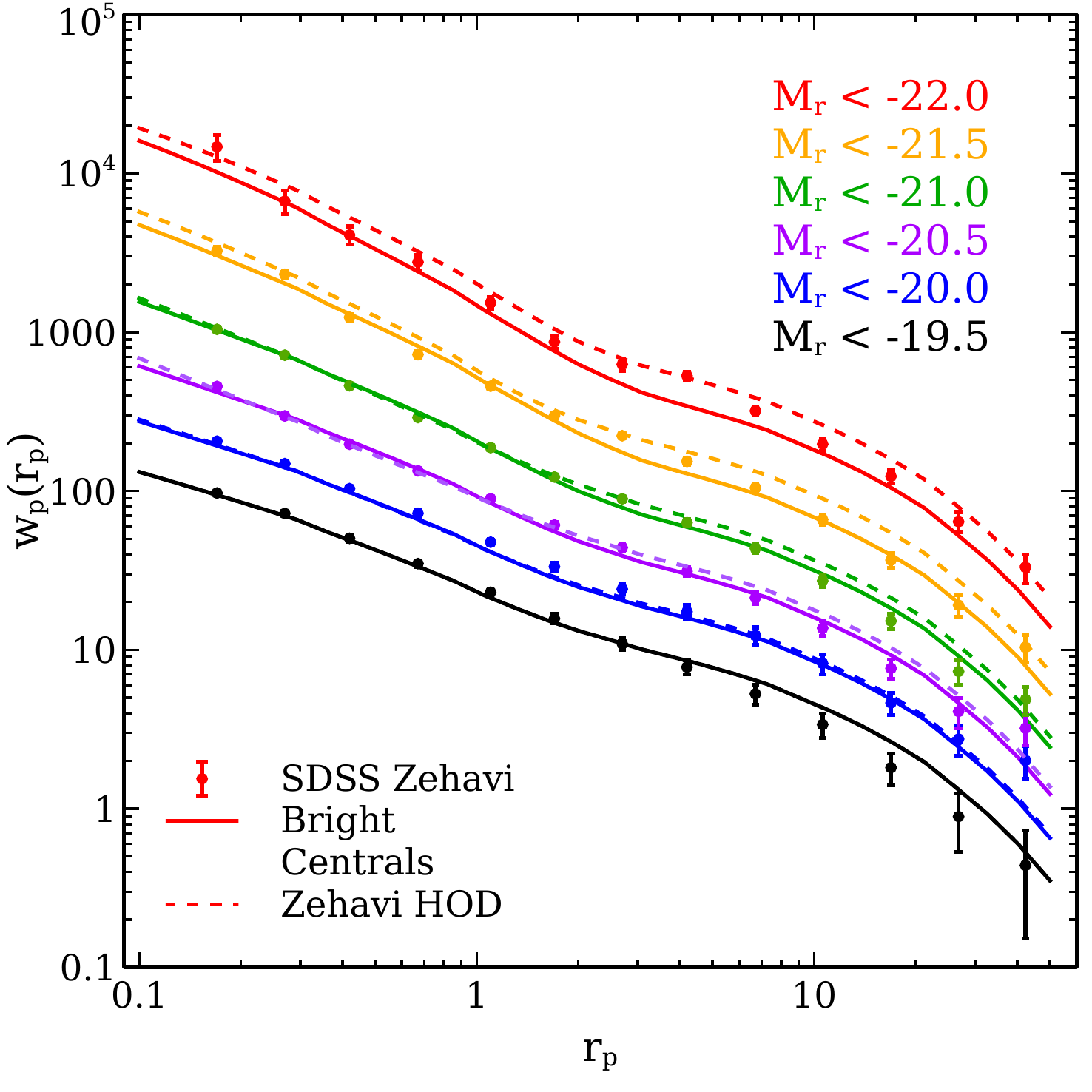}
\caption{Same as Figure~\ref{fig:wp_rp_mass_independent}, except that the solid curves now show the results of the order statistics model in which central luminosities are brightened using \eqn{eq:convolution_g1}.} \label{fig:wp_rp_centrals_convolved}
\end{figure}

Figure~\ref{fig:wp_rp_centrals_convolved} shows the projected correlation function of galaxies at different luminosity thresholds. The solid curves denote the correlation function calculated using our model where the centrals are convolved. It is clear from the Figure that, as with the mass-dependent order statistics model, the new model also agrees well with the SDSS measurements.

\begin{figure}
\centering
\includegraphics[width=0.45\textwidth]{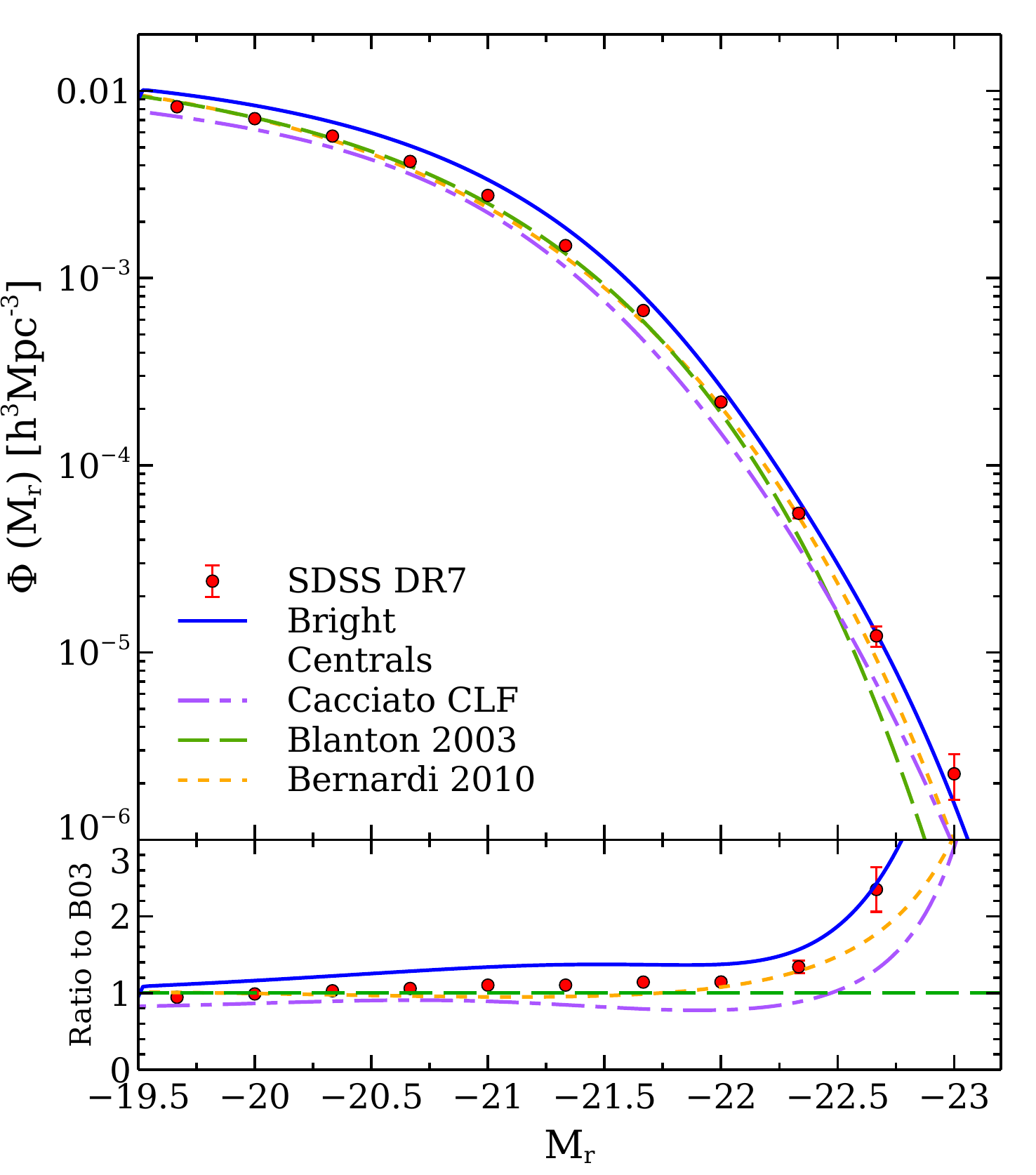}
\caption{Same as Figure~\ref{fig:phi_M_exHOD_mass_dependent}, except that the solid blue curve is now the luminosity function calculated using \eqn{eq:phi_L_m_centrals_convolved} and \eqref{eq:CLF}. For comparison, the purple dot-dashed curve shows the luminosity function calculated using the CLF approach of \citet{Cacciato_2012}.} \label{fig:phi_M_centrals_convolved}
\end{figure}

Finally, the all-galaxy luminosity function can be computed as follows. In this model, the quantity $p_{\rm new}(L|m)$, the probability to find a central of luminosity $L$ in a halo of mass $m$, can be written as
\begin{align}
p_{\rm new}(L|m) &= \frac{p_{\rm cen}(L|m) + \Ns(m) p_{\rm sat}(L|m)}{1+\Ns(m)} \,\, . \label{eq:p_new}
\end{align}
In this expression, $p_{\rm cen}(L|m)$ and $p_{\rm sat}(L|m)$ are the weighted averages of the respective quantities over the number of satellites. Having derived this expression for $p_{\rm new}(L|m)$, we can compute the new CLF $\phi^{\rm new}(L|m)$ as follows:
\begin{align}
\phi^{\rm new}(L|m) &= \fcen(m)\left[1 + \Ns(m)\right]p_{\rm new}(L|m) \notag\\ 
&= \fcen(m)\left[p_{\rm cen}(L|m) + \Ns(m) p_{\rm sat}(L|m)\right]\,, 
\label{eq:phi_L_m_centrals_convolved}
\end{align}
Figure~\ref{fig:phi_M_centrals_convolved} shows the all-galaxy luminosity function calculated using using \eqn{eq:phi_L_m_centrals_convolved} and \eqref{eq:CLF}. We see that the computed luminosity function is in reasonable agreement with the observations.

This order statistics model, modified by brightening the centrals, therefore provides a good description of all the observables we have considered so far. Our formulation is such that the magnitude gap between the brightest and second brightest group galaxy is a \emph{prediction} of the model, as we discuss in the next section.

\section{Magnitude Gap Statistics}
\label{sec:gap}
\noindent
The magnitude gap is defined as the difference between the magnitudes of the brightest and the second brightest galaxy in a group: $\Delta M = M_2 - M_1$, where $M_2$ and $M_1$ are the magnitudes of the second brightest and the brightest galaxy, respectively. The distribution of the magnitude gap is a useful probe of the various physical mechanisms that might affect the evolution of galaxies in groups \citep{Milosavljevic_et_al_2006,vdb+07,ssm07,Hearin_et_al_2012}. In particular, one could characterise the special nature of central galaxies by the difference between the observed distribution of magnitude gaps and that predicted by order statistics \citep{Paranjape_Sheth_2012}. This comparison must be made carefully, using sufficiently large and sufficiently robust data sets \citep{Hearin_et_al_2012,Surhud_2012}. In this section we revisit this comparison using the Yang et al. group catalog to test the predictions of the various models we have discussed above.

\begin{figure*}
\centering
\includegraphics[width=0.45\textwidth]{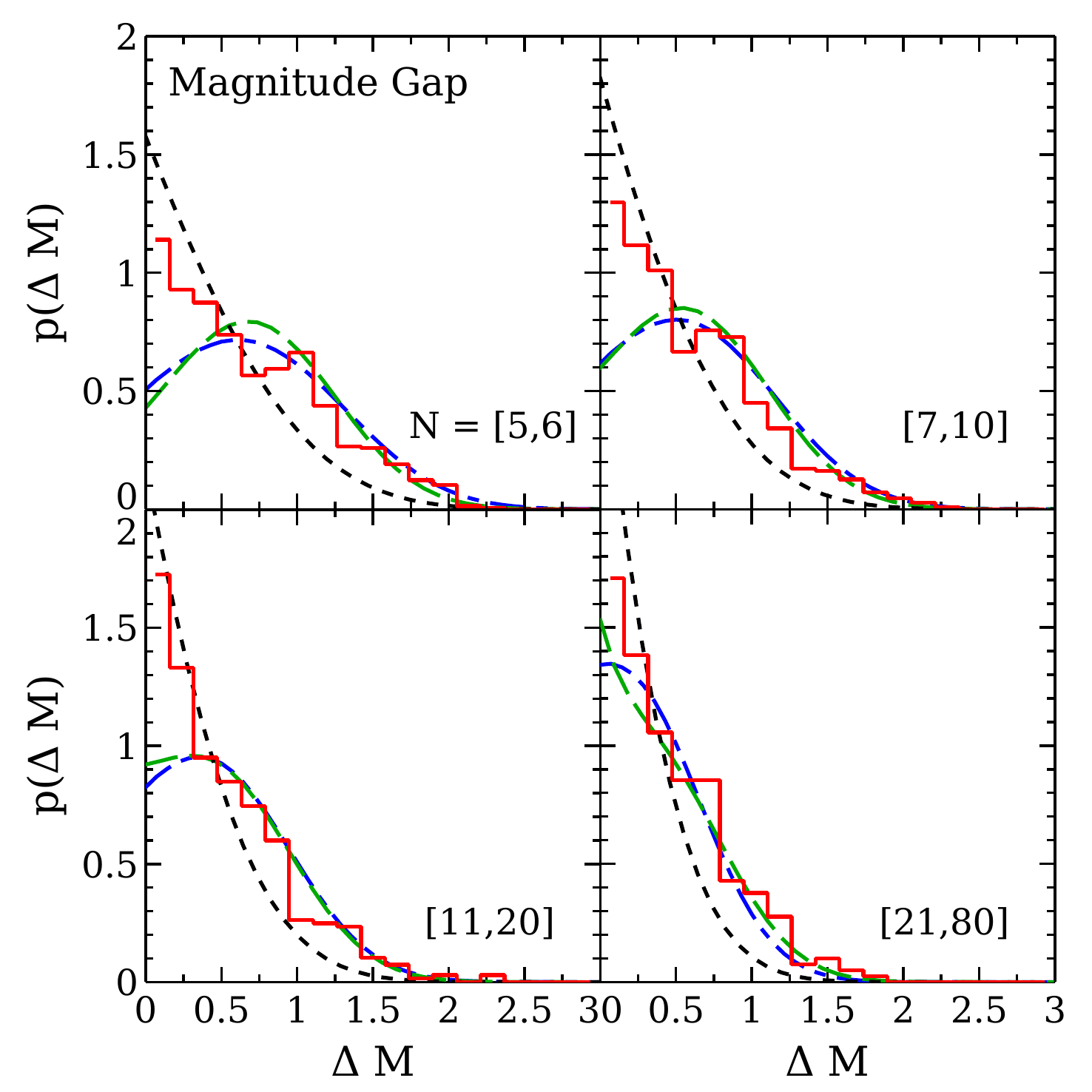}
\includegraphics[width=0.45\textwidth]{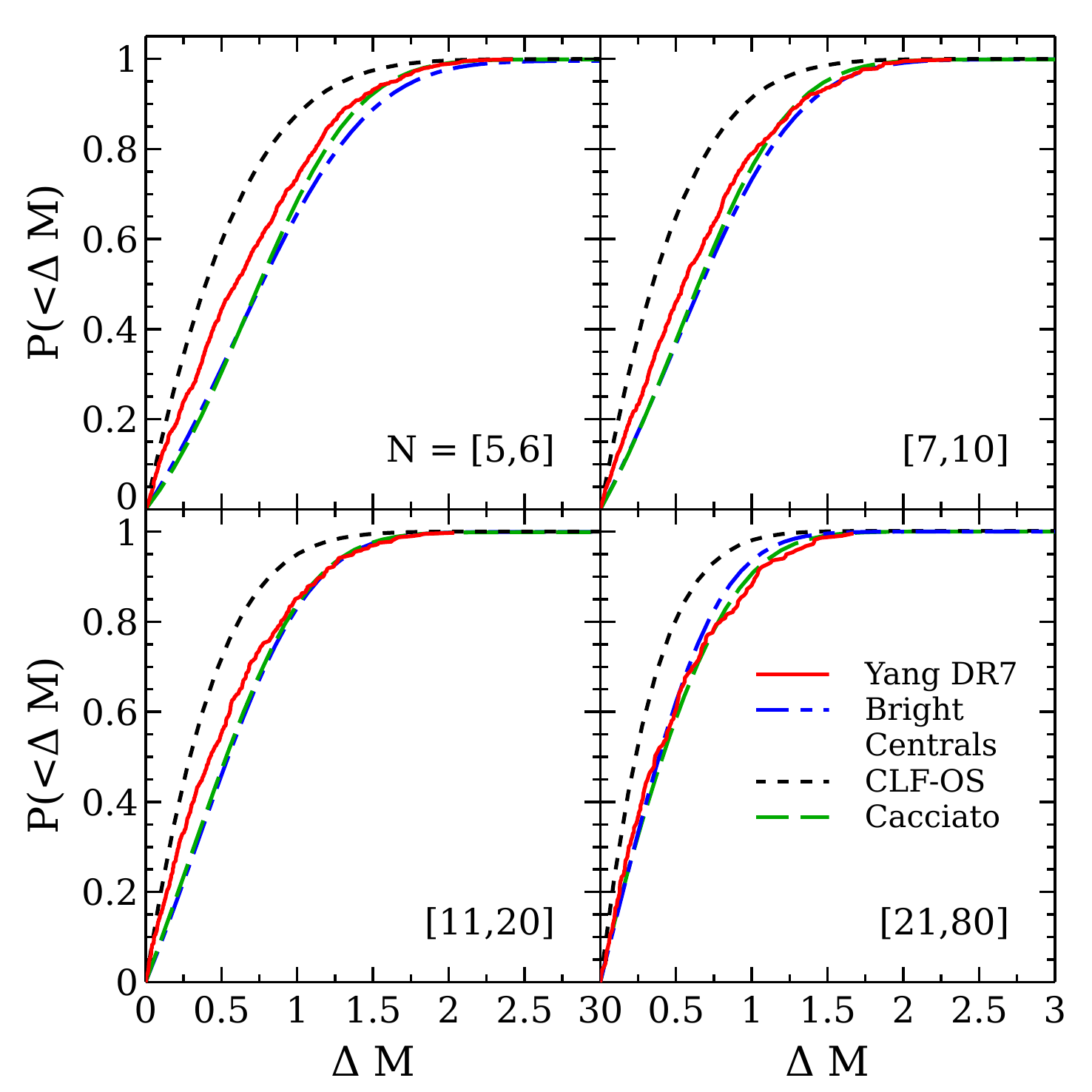}
\caption{Magnitude gap distribution (differential in the \emph{left panel} and cumulative in the \emph{right panel}) in four richness bins. The solid red histograms show the gap directly calculated from Yang et al. catalogue. The  smooth curves show the gap calculated using the CLF approach of \citet[][dashed green]{Cacciato_2012}, the CLF-based order statistics model of section~\ref{sec:extreme_value_mass_dependent} using \eqn{eq:gap_exHOD} (black dotted), and the order statistics model of section~\ref{sec:centrals_convolved} with centrals brightened, using \eqn{eq:gap_CLF_centrals_convolved} (blue dot-dashed).}
\label{fig:magnitude_gap_all}
\end{figure*}

In case of the order statistics model with a universal luminosity function (section~\ref{sec:extreme_value_mass_independent}), the distribution of the magnitude gap in groups containing $N$ galaxies each is given by \citep{Paranjape_Sheth_2012},
\begin{align}
&p(\Delta M |N)\notag\\
&= N(N-1) \int_{-\infty}^{M_{\rm max}} \der M\,p(M)\,p(M-\Delta M)\,p(>M)^{N-2} \,.
\label{eq:gap_exHOD}
\end{align}
For a CLF-based order statistics model (section~\ref{sec:extreme_value_mass_dependent}) one must simply replace the luminosity distribution $p(M)$ and its cumulative distribution $p(>M)$ with $p(M|m)$ and $p(>M|m)$, respectively, in \eqn{eq:gap_exHOD}, hence obtaining the gap distribution at fixed richness \emph{and} halo mass $p(\Delta M |m,N)$. This can then be averaged over halo mass and binned in satellite number (see, e.g.,  equation~\ref{eq:brightness_distribution_binned_version}) to obtain the gap distribution in different group richness bins. 

For the order statistics model with centrals brightened (section~\ref{sec:centrals_convolved}), as well as for the standard CLF approach \citep{Yang_et_al_2003}, the gap distribution is computed differently. In these cases,
\begin{align}
&p(\Delta M|m,N)\notag\\
&=  (N-1) \int_{-\infty}^{M_{\rm max}} \der M\,p_{\rm cen}(M-\Delta M|m,N)\,p_{\rm sat}(M|m,N)\notag \\
&\ph{(N-1)\int\der M p_{cen}(M)}
\times p_{\rm sat}(>M|m,N)^{N-2} \label{eq:gap_CLF_centrals_convolved} \,\, .
\end{align}
Although we have displayed a generic dependence of $p_{\rm cen}$ and $p_{\rm sat}$ on both group richness $N$ and halo mass $m$ in \eqn{eq:gap_CLF_centrals_convolved}, we note that the dependence on $N$ in our model of section~\ref{sec:centrals_convolved} is a consequence of order statistics; in the standard CLF approach, $p_{\rm cen}$ and $p_{\rm sat}$ depend only on halo mass.

Figure~\ref{fig:magnitude_gap_all} shows the distribution of the magnitude gap computed in different models. 
We see that the predictions of the order statistics model with centrals brightened are in excellent agreement with those of the standard CLF approach, whereas the order statistics model without brightening predicts substantially smaller gaps (as expected from the fact that it underpredicts BGG luminosities). None of the models, however, agree with the measurements from the Yang et al. catalogue, except in the highest richness bin. 
In making this comparison, however, one must keep in mind that the measured gap distribution is sensitive to errors in classifying galaxies as centrals and satellites. In fact, recent work by \citet{Campbell_et_al_2015} has shown that the algorithm of \citet{Yang_et_al_2007} leads to misclassification of centrals as satellites and vice-versa, at the level of $20$-$30$ per cent. A detailed study of the effect of these misclassifications on the gap distribution is beyond the scope of the present work. 

Finally, an alternative way of framing the comparison between the observed and modelled magnitude gap distribution is to ask whether this distribution depends on \emph{both} group richness $N$ and halo mass $m$ in a non-trivial way. \citet{Hearin_et_al_2012} argued that the observed gap distribution in the \citet{Berlind_et_al_2006} group catalog (updated to SDSS DR7) does, in fact, show a dependence on halo mass at fixed richness. 
This was further elaborated on by \citet{Surhud_2012}, who also calculated the distributions $p(\Delta M|m,N)$ using the standard CLF approach. 
The order statistics model based on a universal luminosity function \citep{Paranjape_Sheth_2012} does not predict such a dependence, while the standard CLF model, as well as the models discussed in sections~\ref{sec:extreme_value_mass_dependent} and~\ref{sec:centrals_convolved}, do. It is therefore interesting to compare the predictions of these models with the results mentioned above.

Figure~\ref{fig:p(DM|m,N)} shows the distribution $p(\Delta M|m,N)$ for $N=21$ and three choices of halo mass near the characteristic mass scale.
The dashed curves show the standard CLF prediction and the solid curves show the prediction of our order statistics model with centrals brightened. Clearly, both models show a substantial mass dependence, with the various distributions being very different from the mass-independent prediction of the universal order statistics model, shown as the dotted curve. (The predictions of the CLF-based order statistics model \emph{without} brightening the centrals have a much weaker mass dependence and lie close to the universal model; to avoid clutter we have chosen not to display these.) 

\begin{figure}
\centering
\includegraphics[width=0.45\textwidth]{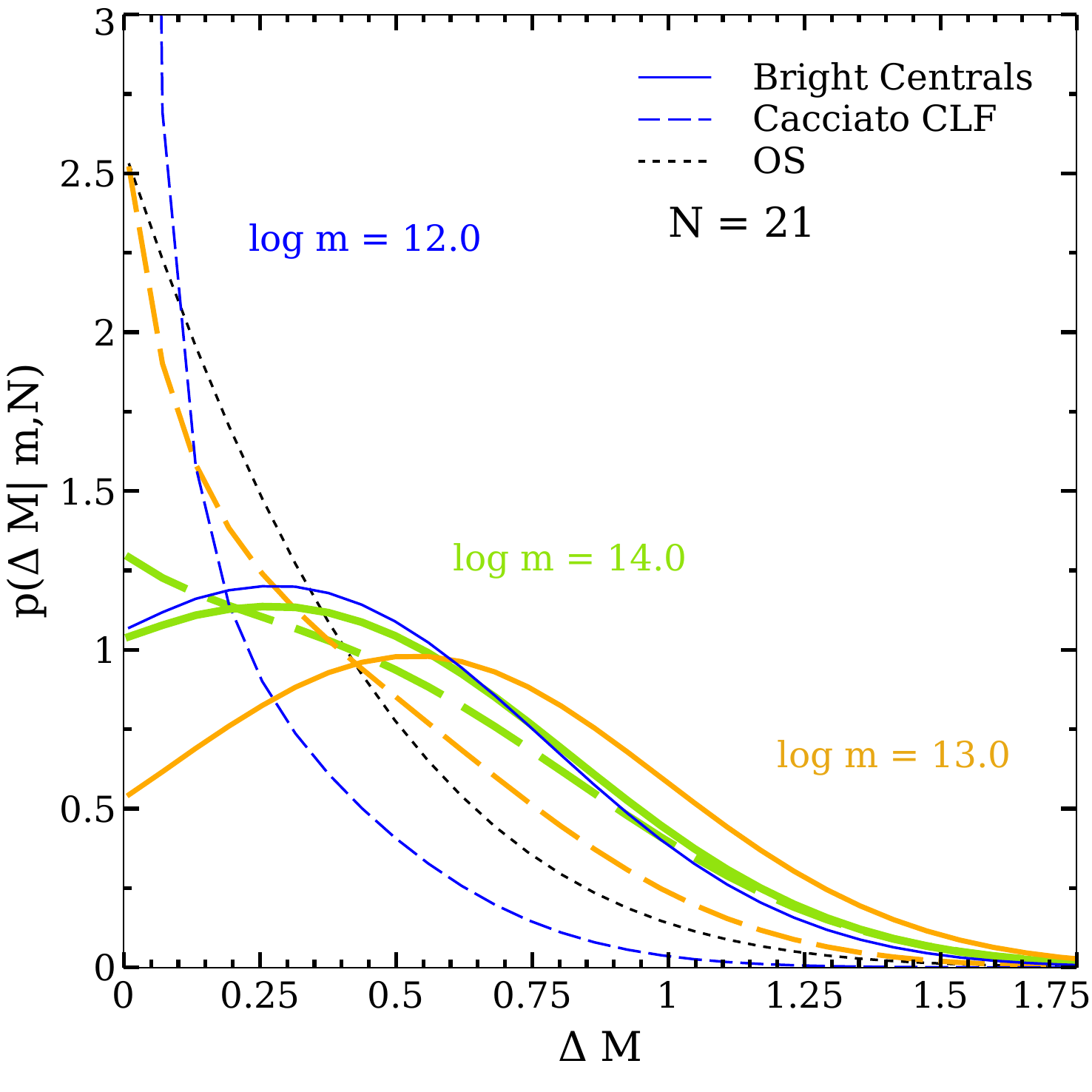}
\caption{Distribution of magnitude gap $p(\Delta M|m,N)$ at fixed group richness $N=21$  and for three values of mass: in increasing order of line thickness we show $\log(m)=12.0,13.0,14.0$ (blue, yellow, green). The solid curves show the results of the order statistics model with centrals brightened (equation~\ref{eq:gap_CLF_centrals_convolved}) and the dashed curves show the standard CLF results from \citet{Surhud_2012}. For comparison, the dotted black curve shows the mass-independent prediction at this group richness of the universal order statistics prediction \citep[][see also equation~\ref{eq:gap_exHOD}]{Paranjape_Sheth_2012}. See text for a discussion.} 
\label{fig:p(DM|m,N)}
\end{figure}

What is interesting, however, is the fact that the trend with mass in our model is very different from that in the standard CLF model. Whereas the CLF model predicts a monotonically increasing median and width of the magnitude gap as a function of mass, our model predicts a median that first increases until $\log(m)=13.0$ and then decreases. This can be traced back to our choice of brightening $\mu$ in \eqn{eq:expression_mu} which has a maximum at this mass scale and falls to zero on either side. Since the satellite luminosities in our model are not modified from the order statistics form, at both large and small masses our model predicts a gap distribution closer to the universal model, while around $\log(m)=13.0$ (corresponding to maximum brightening of the central), the difference from the universal model is the largest. 

Although the detailed shapes of the gap distributions depend on our particular choice of brightening kernel (which was chosen only as a toy model), we believe that the non-monotonicity seen above is a robust feature of such models, corresponding to the fact that the brightening relative to order statistics is only required around $\log(m)\simeq13$. 
One might worry, however, that this mass scale, as well as the amount of brightening, may depend on our specific choice of galaxy sample. To test whether our prescription is robust, we repeated our analysis for a subset of the sample discussed above, keeping only galaxies with $M_r < -20.0$ in the red box in Figure~\ref{fig:SDSS_data}. Excluding faint objects leads to a systematic decrease in the richness of individual groups. Apart from this, however, the distribution of BGG luminosity or magnitude gap for large enough groups must remain unaffected. To see whether our model tracks these effects correctly, we recomputed all the statistics discussed previously, using the new luminosity threshold of $M_{r,{\rm max}}=-20.0$ but keeping \emph{all} other parameters fixed to their original values. We find that the model agrees with the new measurements at the same level as before; Figure~\ref{fig:BGGgapMr20} in the Appendix demonstrates this for the BGG and magnitude gap distributions. Further, we have also found that varying the brightening mass scale by more than about $0.5$dex around its default value of $\log(m)\simeq13$ leads to substantial disagreement with the BGG luminosity distribution and/or the clustering measurements, for both $M_{r,{\rm max}}=-19.5$ and $M_{r,{\rm max}}=-20.0$. Thus, we expect the brightening mass scale to be robust.

Since $\log(m)=13$ is suggestively close to the characteristic halo mass scale at $z=0$, it will be very interesting to explore physical arguments to motivate the choice of the brightening kernel, as well as observational tests of this non-monotonicity. We will return to these issues in future work.

\section{Conclusions}
\label{sec:conclude}
\noindent
We have used the Halo Model framework to explore the implications of assuming that galaxy luminosities obey order statistics. While the simplest of this class of order statistics models is ruled out by observations, we argued that this new framework provides considerable conceptual advantages over the standard approach which treats central galaxies as special from the outset.

Our main conclusions are the following:
\begin{itemize}
\item The assumption that central and satellite luminosities in a group obey order statistics with a universal luminosity distribution $p(L)$ (with the central being the brightest), considerably simplifies the Halo Model framework for galaxy clustering and correctly predicts observed qualitative trends such as an approximately Lognormal distribution of central galaxy luminosities, with a mean (width) that increases (decreases) with halo mass or group richness, as well as a tight correlation between the respective mass scales describing the central and satellite luminosity functions (section~\ref{sec:extreme_value_mass_independent}). However, this model fails to describe the observed luminosity dependence of clustering in SDSS \citep[see also][]{Paranjape_Sheth_2012}. In particular, the model predicts \emph{no} luminosity dependence of large scale clustering.
\item An extension of this model to allow a halo mass dependence $p(L|m)$ in the underlying luminosity function leads to substantial improvement in the comparison with observations of clustering (section~\ref{sec:extreme_value_mass_dependent}). This CLF-based order statistics model also correctly describes the observed  all-galaxy luminosity function, as well as the luminosities of the brightest satellites, but systematically predicts fainter centrals than observed in the SDSS-based group catalogue of \citet{Yang_et_al_2007}.
\item This brings into focus the idea that central galaxies constitute a distinct population that is presumably affected by different physical processes than are the satellites. The effect of these physical processes can be captured by statistically brightening the centrals, over and above the order statistics prediction for their luminosities \citep[section~\ref{sec:centrals_convolved}; see also][]{Shen_et_al_2014}. The resulting model is in good agreement with all the observables mentioned above.
\item The magnitude gap between the brightest and second brightest group galaxy is then a \emph{prediction} of this model, and agrees reasonably well with measurements in the Yang et al. catalogue (section~\ref{sec:gap}). Our model also predicts a halo mass dependence of the gap distribution at fixed group richness: $p(\Delta M|m,N)\neq p(\Delta M|N)$. In contrast to the standard CLF approach where this mass dependence is monotonic for the median gap, our model predicts a \emph{non-monotonicity} with halo mass (Figure~\ref{fig:p(DM|m,N)}) which will be interesting to test observationally.
\end{itemize}

Our analysis above has several potential applications. For example, the simple order statistics model based on a universal luminosity function can be used to set well-motivated priors when fitting HOD parameters. Our formulation of the two CLF-based order statistics models can also be easily adapted to a Markov Chain Monte Carlo approach for determining the various free parameters, exactly as in the standard CLF approach. In parallel, the additional brightening of central luminosities over and above the order statistics prediction, modelled using \eqn{eq:convolution_g1}, can provide a useful language for comparing the statistically motivated Halo Model with more physically driven models based on semi-analytical calculations and numerical simulations. The fact that the data seem to force us to preferentially brighten BGGs in haloes of mass $m\sim10^{13}h^{-1}\Msun$ close to the characteristic mass at $z=0$, is particularly striking in this regard. 
Finally, we are also aware that the luminosity statistics studied in this work, particularly the magnitude gap, may be sensitive to recalibrations of SDSS luminosity measurements, since these are most important for bright galaxies in cluster environments \citep[see, e.g.,][]{Bernardi_et_al_2010,Bernardi_et_al_2013,Meert_et_al_2015}.  These recalibrations are known to affect the determination of the HOD \citep{Shankar_et_al_2014}.  However, if the recalibration preserves the rank ordering of galaxies and simply boils down to an overall shift in log(Luminosity), then our results will not be affected.  We leave a fuller exploration of this, and the other ideas above, to future work.

\section*{Acknowledgements}
\noindent
We thank Surhud More for useful discussions. 
We are grateful to the SDSS collaboration for releasing their data set, and to Yang et al. for making their group catalog publicly available.
NP acknowledges  the financial support from the Council of Scientific and Industrial Research (CSIR), India as a Shyama Prasad Mukherjee Junior Research Fellow. The research of AP is supported by the Associateship Scheme of ICTP, Trieste and the Ramanujan Fellowship awarded by the Department of Science and Technology, Government of India.


\bibliography{exHOD_bibliography} 



\appendix

\section{Robustness of brightening prescription}
\label{app:bright}

\begin{figure*}
\centering
\includegraphics[width=0.45\textwidth]{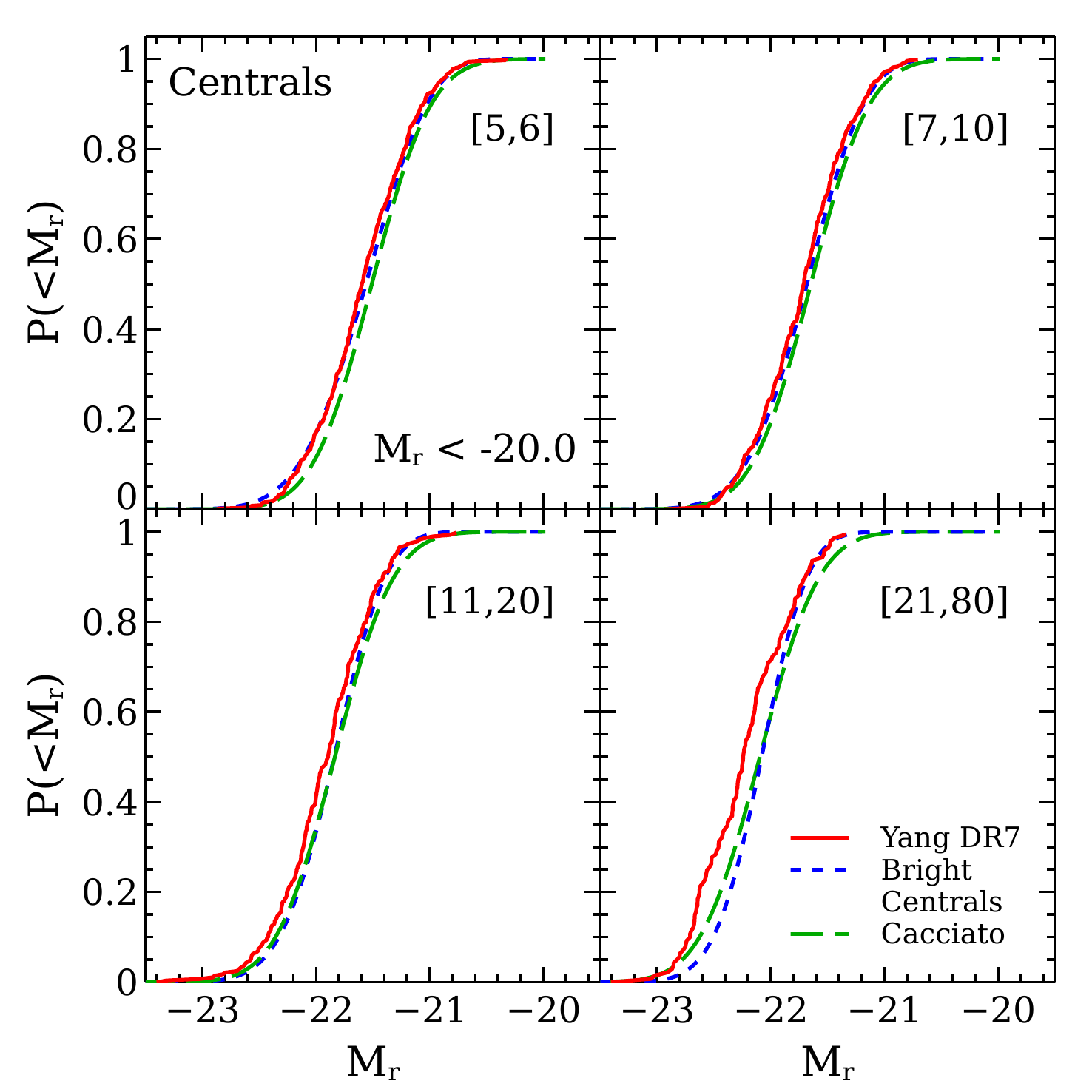}
\includegraphics[width=0.45\textwidth]{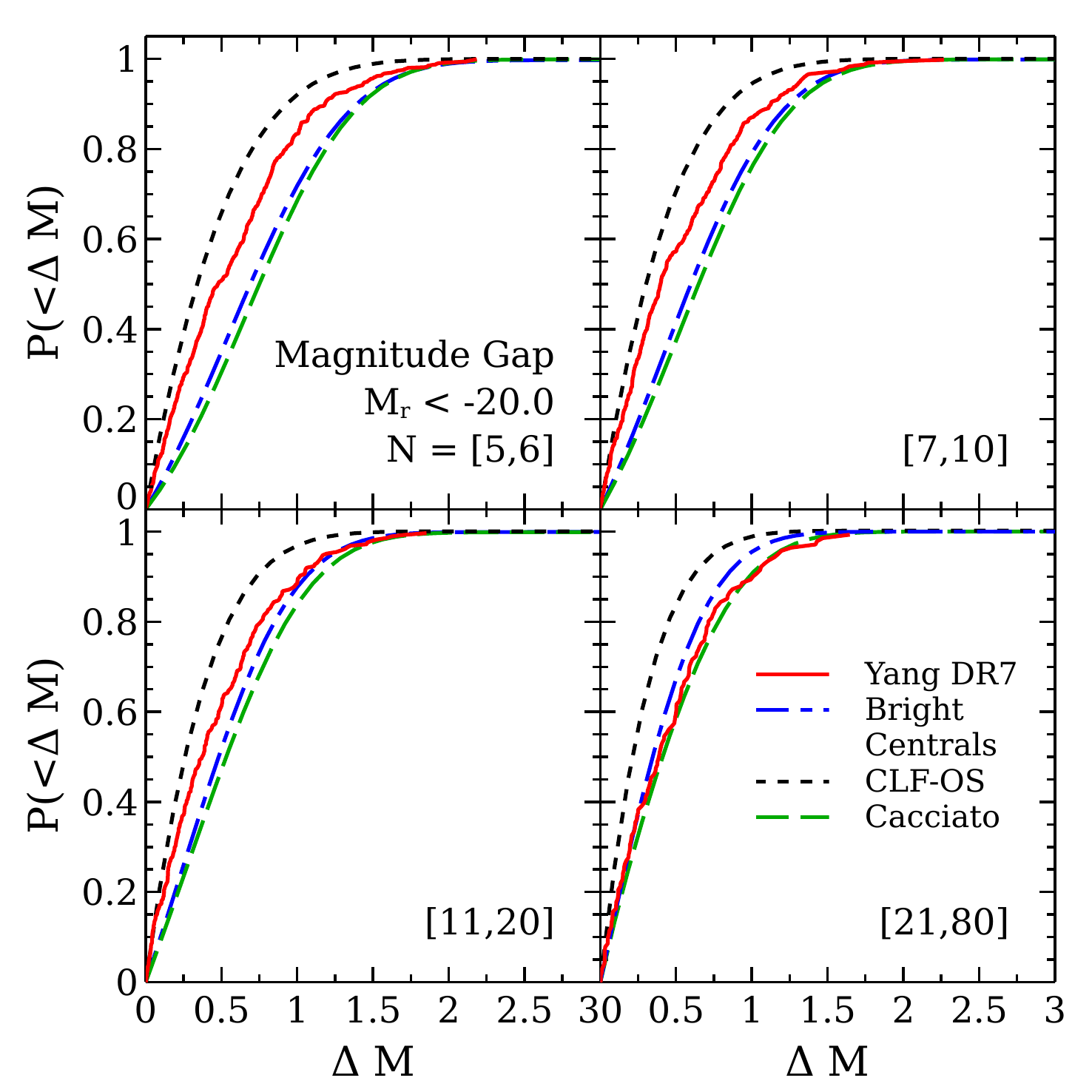}
\caption{Cumulative distributions of the BGG luminosity \emph{(left panel)} and magnitude gap \emph{(right panel)}, for the SDSS sample defined by luminosity threshold $M_{r,{\rm max}}=-20.0$ and in the same volume as defined by the red box in Figure~\ref{fig:SDSS_data}. The smooth curves show the corresponding analytical results as indicated. The left and right panels may be compared, respectively, with the right panels of Figures~\ref{fig:BGG_centrals_convolved} and~\ref{fig:magnitude_gap_all}. The formatting is identical to those Figures.}
\label{fig:BGGgapMr20}
\end{figure*}

\noindent
To test the robustness of our order statistics model with centrals brightened, particularly that of the choice of  brightening kernel \eqref{eq:kernel}, we recomputed all the statistics discussed in the main text, using the luminosity threshold of $M_{r,{\rm max}}=-20.0$ but keeping \emph{all} other parameters fixed to their original values. We compared these with corresponding measurements in a subset of the SDSS DR7 sample described in the text, by discarding galaxies fainter than this new threshold, while keeping the volume fixed to that defined by the red box in Figure~\ref{fig:SDSS_data}.

We have found that the model agrees with the new measurements at the same level as before. Figure~\ref{fig:BGGgapMr20} shows the analytical and measured BGG and magnitude gap distributions for the new sample; these may be compared with the results for the fainter sample in Figures~\ref{fig:BGG_centrals_convolved} and~\ref{fig:magnitude_gap_all}. While there are small differences with respect to the earlier comparison for groups with few member galaxies, at the high richness end our model continues to provide a good description of the measurements. This is discussed further in the main text.


\bsp	
\label{lastpage}
\end{document}